\documentclass[12pt,preprint]{aastex}
\usepackage{epsfig}
\usepackage{multirow}
\usepackage{longtable}
\usepackage{natbib}
\usepackage{amsbsy}
\usepackage{rotating}
\newcommand{\agile}{\textit{Agile}}
\newcommand{\rfilter}{$r$--band}

\newcommand{\mearth}{M$_\earth$}

\newcommand{\ie}{\textit{i.e.}}
\newcounter{parnum}
\newcommand{\N}{%
   \noindent\refstepcounter{parnum}%
    \makebox[\parindent][l]{\textbf{\Roman{parnum}.}}}
\begin{document}
\title{APOSTLE Observations of GJ 1214b: System Parameters and Evidence for Stellar Activity}
\author{
  P.~Kundurthy\altaffilmark{1},
  E.~Agol\altaffilmark{1},
  A.C.~Becker\altaffilmark{1},
  R.~Barnes\altaffilmark{1,2},
  B.~Williams\altaffilmark{1},
  A.~Mukadam\altaffilmark{1}
}
\altaffiltext{1}{Astronomy Department, University of Washington, Seattle, WA 98195}
\altaffiltext{2}{Virtual Planetary Laboratory, USA}

\begin{abstract}
We present three transits of GJ 1214b, observed as part of the Apache Point Observatory Survey of Transit Lightcurves of Exoplanets (APOSTLE). We used APOSTLE \rfilter\ lightcurves in conjunction with previously gathered data of GJ 1214b to re-derive system parameters. By using parameters such as transit duration and ingress/egress length we are able to reduce the degeneracies between parameters in the fitted transit model, which is a preferred condition for Markov Chain Monte Carlo techniques typically used to quantify uncertainties in measured parameters. The joint analysis of this multi-wavelength dataset confirms earlier estimates of system parameters including planetary orbital period, the planet-to-star radius ratio and stellar density. Estimating the absolute mass and radius of the planet directly depend on how various stellar parameters are derived. We fit the photometric spectral-energy distribution of GJ 1214 to derive stellar luminosity and estimated its absolute mass and radius from known mass-luminosity relations for low-mass stars. From these derived stellar properties and previously published radial velocity data we were able to refine estimates of the absolute parameters for the planet GJ 1214b. Transit times derived from our study show no evidence for strong transit timing variations. Some lightcurves we present show features that we believe are due to stellar activity. During the first night we observed a 0.8\% rise in the out-of-eclipse flux of the host star lasting approximately 3 minutes. The trend has a characteristic fast-rise exponential decay shape commonly associated with stellar flares. During the second night we observed a minor brightening during the transit. Due to its symmetric shape we believe this feature might have been caused by the planet obscuring a star-spot on the stellar disk.
\end{abstract}
\keywords{}

\section{Introduction}
\label{sec-intro}
Exoplanets detected using both transits and the radial velocity (RV) technique offer the unique opportunity to measure many of their physical properties. Transits can place limits on the radius of an exoplanet given the radius of the star, and transits can also constrain orbital parameters such as inclination and orbital period. When coupled with RV measurements, we can constrain the mass ($M_{p}$), and the average density of a planet ($\rho_{p}$). We may therefore probe the interiors of transiting exoplanets and constrain bulk composition and formation models \citep{fortney08,torres08,baraffe08}. Here we present and interpret three lightcurves of the transiting planet GJ 1214b \citep{charbonneau09}, a planet unlike any in our Solar System.

Prior to 2009, all transiting exoplanets were found with radii consistent with the giant planets in our Solar System, \ie\ they had gaseous envelopes. As surveys improved, they began to reach sensitivities which could detect smaller rocky planets. In our Solar System, Neptune has a mass of 17 \mearth while the Earth is the largest terrestrial planet, suggesting that the transition mass between rocky and gaseous planets lies somewhere between these values. With no such ``transition'' object in the Solar System, we must rely on theory, which predicts that $\sim 10$ \mearth\ is the critical mass \citep{pollack96}, although it could be as low as 2 \mearth \citep{ikoma01} or as high as 16 \mearth \citep{lissauer09}. The 10 \mearth\ limit should therefore be seen as a sort of median of theoretical results, and not a true boundary between rocky and gaseous worlds. The discovery of transiting planets between 1 and 17 \mearth therefore provides critical insight into the planet formation process.

Two transiting planets are now known with 1\mearth $\le M_{p} \le$ 10 \mearth; CoRoT-7b \citep{leger09,queloz09}, and GJ 1214b \citet{charbonneau09}. The former appears to be rocky \citep{leger09}, while the latter may contain significant amounts of water or a gaseous envelope \citep{rogersseager10}. In order to address the ambiguities in the planet formation process, the radii of these planets must be measured accurately and precisely. The CoRoT satellite already surpasses all other projects in the ability to make follow-up observations of CoRoT-7b (153 transits reported by \citet{leger09}). The relatively recent detection and hence the dearth of follow-up measurements on GJ 1214b led us to focus on this object; a Super-Earth planet orbiting M dwarf star 13 pc from Earth \citep{charbonneau09}. The planetary nature of this transit has been confirmed by \citet{sada10}.

The discovery of such a small-sized planet bodes well for transit searches for habitable planets around M-Dwarfs. The low luminosity of M-Dwarfs mean their habitable zones are very close to the star \citep{kasting93,selsis07}. This increases the transit probability \citep{boruckisummers84}, and hence some of the first planets to be characterized as rocky and in the habitable zone may well be transiting planets around M-Dwarfs. Although GJ 1214b most likely possesses a Hydrogen-rich envelope, the detection of small planets such as this heralds the discovery of rocky habitable worlds if they exist. M-Dwarfs make up a very large fraction of the stellar component of the Milky Way \citep{MS79,reid02}, so the prospect of habitable planets around M-Dwarfs raises the very interesting possibility that life-bearing planets may be fairly common in the galaxy.

In this paper we report observations of three transits of GJ 1214b in 2010. In $\S$ \ref{sec-data} we outline our observations and data reduction techniques; in $\S$ \ref{sec-syspars} we describe our lightcurve model and the use of Markov Chain Monte Carlo (MCMC) techniques to constrain system parameters from single and multi-wavelength data. In $\S$ \ref{sec-ttv} we report the absence of strong transit timing variations (TTV). In $\S$ \ref{sec-mr} we describe the derivation of various stellar and planetary parameters for the GJ 1214 system. In $\S$ \ref{sec-stellaractivity} we discuss the detection of a flare and a possible spot-crossing event.  Finally, in $\S$ \ref{sec-conclusions} we summarize our findings.

\section{Data}
\label{sec-data}

\subsection{APOSTLE Observations}
\label{sec-obs}
We observed three transits of GJ 1214b on UT dates 2010-04-21, 2010-06-06 and 2010-07-06 using the ARC 3.5m Telescope at Apache Point, New Mexico. All observations were made with \agile, a high-speed time-series CCD photometer based on the design of \textit{Argos} \citep{nathermukadam04}. \agile\ is a charge transfer CCD, that collects photons from the target at a 100\% duty cycle. During transit observations the charge on \agile\ was read out at 45 sec intervals using GPS-synchronized pulses with an absolute timing accuracy of less than a millisecond. Observations were made using \agile's medium gain, slow readout mode with frames binned by a factor of 2, yielding a plate-scale of 0.258 $\arcsec$/pixel. All APOSTLE lightcurves presented in this work were collected using the $r$--filter, which is similar to the SDSS $r$ filter \citep{fukugita96}.

During observations we defocused the telescope to spread the stellar point-spread function (PSF) across multiple pixels in order to minimize systematics introduced by improper flat-fielding and to increase the integrated counts from the star. Count rate stability is affected by variable conditions during observing. We monitored the count rate and adjusted the telescope focus by small increments to raise or lower the maximum counts on the brightest star. By adjusting the focus we kept the maximum counts between 40k and 55k ADU. We did this because the instrument is known to show a non-linear response at count-levels greater than $\sim$55k, and saturates at 61k counts. During stable conditions these adjustments were made every 10 to 20 minutes while during poor conditions the adjustments were more frequent, at roughly 5 minute intervals. Less than 7$\%$ of frames on a given night were lost to saturation. For all our observations the comparison star USNO-B1.0 0949-0280051 was the brightest object within \agile's field-of-view, 95$\arcsec$ to the north of GJ 1214 \citep{monet03}. This object was brighter than GJ 1214 by a factor of 1.4 in the \rfilter. It was also the only star brighter than GJ 1214 in \agile's field-of-view, and hence was the only one used for differential photometry. On each night we also collected twilight sky flats and dark frames (which were also used to correct for bias).

\subsection{Image Reductions}
\label{sec-imred}
We used a customized data reduction pipeline, written in Interactive Data Language (IDL) to process \agile\ data. It performs standard image processing steps like dark subtraction and flat fielding, but also implements non-linearity corrections unique to \agile. The pipeline also creates an uncertainty map of the processed images by propagating pixel-to-pixel errors through each step of the reduction. In addition to the Poisson photon counting errors and read-noise from the science images, the pipeline propagates the variance on the master dark and master flat during the reduction. Errors were also propagated for those pixels where the counts exceeded the non-linearity threshold ($\sim$55k counts) using the uncertainties in the empirically derived non-linearity correction function. Typically 1-5$\%$ of frames went above this threshold. The correction factor for counts on non-linear pixels was typically 1.45$\%$ or less.

\subsection{Photometry}
\label{sec-phot}
We used SExtractor \citep{bertinarnouts96} to derive initial centroids of our defocused stars. Coordinates obtained from SExtractor were then used for circular aperture photometry with the PHOT task in IRAF's NOAO.DIGIPHOT.APPHOT package. We derived flux estimates from a range of circular apertures with radii between 5-50 pixels, at intervals of 1 pixel. An outlier--rejected global median was used as the sky estimate. An optimal aperture was selected where the RMS in the out-of-eclipse lightcurve was minimized. For GJ 1214 data this aperture was typically 15-16 pixels in radius. This size was roughly 4 times the Half-Width at Half Maximum (HWHM) of the PSFs. To derive photometric errors we extracted counts from the error frames using the same centroids and apertures used for photometry on the target frames. Estimating photometric errors in this manner yielded uncertainties that were greater (by 50\%) than the default errors reported by PHOT. It is useful to note that this method takes into account sources of error which are otherwise ignored by standard photometric techniques, e.g. fluctuations in the flat field and dark frame, and is thus more thorough. Photometric precision for our GJ1214 lightcurves was typically $\sim 0.001$ magnitudes from 45sec exposures.

\subsection{Time Coordinates}
\label{sec-timecoor}
The IAU recommended time coordinate for the proper analysis of event timing data is TCB \citep[Barycentric Coordinate Time,][]{kaplan05}. There are two primary advantages of using this time coordinate: 1) TCB is the best approximation we have to taking measurements from an inertial reference frame. In this system, the time-of-arrival of photons are described as measured by a clock at the solar system barycenter (SSB) assuming gravitational effects on light travel time due the largest solar system bodies are removed \citep{edwards06}. 2) TCB is based on the SI definition of the time unit -- $second$ \citep{standish98}.

We followed the prescription described in \citet{seidelmannfukushima92}, \citet{standish98}, \citet{hobbs06}, and \citet{edwards06} to transform the mid-exposure UTC time (Coordinated Universal Time) into TCB:
\begin{equation}
\label{eqn-tcb}
TCB (JD) = UTC (JD) + C_{cor} + G_{cor} + E_{cor},
\end{equation}
where TCB (JD) and UTC (JD) are the Julian Day representations of TCB and UTC respectively; $C_{cor}$ is the clock correction term (used to convert UTC to Terrestrial Time (TT)), $G_{cor}$ is the geometric light travel time correction (or Roemer delay), and $E_{cor}$ is relativistic correction (the Einstein delay) \citep[see also,][]{hobbs06,edwards06,eastman10}. The method is also described in detail by \citet{eastman10} with reference to exoplanetary transits. However, \citet{eastman10} recommend the closely related TDB (Barycentric Dynamical Time) system instead of TCB for reporting timing measurements. The difference between these two coordinates is very slight. However, \citet{standish98} discusses how the TDB system does not represent a physical time measure but is related to TCB by a constant offset and scaling factor. Due to this the times represented by TDB are in units that are subtly different from the widely used SI second \citep{hobbs06}. UTC and Terrestrial Times (TT) reported by observatory and GPS (Global Positioning System) clocks are typically in the SI system.

\subsection{Other lightcurves}
\label{sec-mearthdata}
The MEarth team graciously shared the lightcurves presented in \citet{charbonneau09}, which were either from the array of MEarth (8$\times$40cm) telescopes or $z$-filter observations from the Fred Lawrence Whipple Observatory's 1.2m telescope (FLWO1.2m). We included these data in our multi-wavelength analysis of system parameters. We modified the time coordinate to the TCB$_{JD}$ system as described in $\S$\ref{sec-timecoor}, and when data on nuisance parameters were available, we included them in our detrending analysis.

\section{System Parameters}
\label{sec-syspars}

\subsection{Transit Lightcurve Model and Detrending}
\label{sec-model-det}
APOSTLE transit lightcurves of GJ 1214b are shown in Figure \ref{figure_lcs}. We use the lightcurve models described in \citet{mandelagol02} assuming quadratic stellar limb-darkening and a circular planetary orbit. Our transit code can fit for multiple transits simultaneously and allows for multiple sets of limb-darkening coefficients when data gathered using different filters are used. For the analysis presented in this paper we used models which fit parameters in either one of the following two sets: for lightcurves from a single-filter, we used \textit{Set 1:} $\boldsymbol{\theta}_{1}$ = \{$t_{T},t_{G},D,v_1,v_2,T0_{i...N_T}$\} and for the joint analysis of lightcurves gathered at different wavelengths, we used \textit{Set 2:} $\boldsymbol{\theta}_{2}$ = \{$t_{T},t_{G}, R^{2}_{p}/R^{2}_{\star}, v_{1,j...N_F}, v_{2,j...N_F}, T0_{i...N_T} $\}.

The parameters $t_{T}$ and $t_{G}$ are approximately the transit duration and ingress/egress duration respectively \citep[same as $T$ and $\tau$ from][]{carter08}. The transit duration is defined as the time between the middle of ingress to the middle of egress. The ingress duration describes the time between the start of the eclipse till the planet has completely crossed the limb of the star. The duration of egress is assumed to be the same. In the limit of zero eccentricity orbits these two parameters can be approximated as,
\begin{equation}
\label{eq-tt}
t_T = 2 \frac{R_{\star}}{v} \sqrt{1-b^2}
\end{equation}
\begin{equation}
\label{eq-tg}
t_G = 2 \frac{R_{p}}{v \sqrt{1-b^2}}
\end{equation}
where the orbital speed $v = 2\pi a/P$ and the impact parameter $b = a\sin{i}/R_{\star}$. The terms $a$, $P$, $i$, $R_{p}$ and $R_{\star}$ are the semi-major axis, orbital period, inclination, planetary radius and stellar radius respectively \citep{carter08}.

We assume quadratic limb-darkening as described in \citet{mandelagol02}, and assume the transit depth approximately changes as,
\begin{equation}
D(b) = \frac{R^{2}_{p}}{R^{2}_{\star}} (1- \gamma_1 (1-\sqrt{1-b^2})-\gamma_2 (1-\sqrt{1-b^2})^2)
\end{equation}
here $\gamma_1$ and $\gamma_2$ are the quadratic limb-darkening coefficients from \citet{mandelagol02} and $R^{2}_{p}/R^{2}_{\star}$ is the square of the planet-to-star radius ratio. Solving for the impact parameter in equations \ref{eq-tt} and \ref{eq-tg} gives $b = \sqrt{1 -(t_{T}/t_{G})(R_p/R_{\star})}$. So given $t_T$, $t_G$ and the limb-darkening coefficients, the maximum depth at mid-transit $(D)$ can be determined by solving a sextic equation in $R^{2}_{p}/R^{2}_{\star}$. The difference between the two fit parameter sets ($\theta_{1}$ and $\theta_{2}$) are the variables $D$ and $R^{2}_{p}/R^{2}_{\star}$. Since the limb-darkening coefficients are dependent on the waveband used for observations, $R^{2}_{p}/R^{2}_{\star}$ is commonly used to fit for the depth of transit lightcurves, but $D$ might be better suited for constraining the transit-depth of single-filter data, especially when the star is expected to be strongly limb-darkened.

The terms $v_1$ and $v_2$ are linear combinations of the quadratic limb-darkening coefficients; $v_1 = \gamma_1+\gamma_2$ and $v_2 = \gamma_1-\gamma_2$. These linear combinations were used since it is known that directly fitting for limb-darkening coefficients results in strongly anti-correlated error distributions for transit parameters \citep{brown01}. In order to avoid unphysical limb-darkening profiles we applied the bounds $v_1 + v_2 > 0$ and $0 < v_1 < 1$. The $T0$ terms are the times of transit center. The subscripts ${i...N_T}$ and ${j...N_F}$ are used to denote multiple transits ($N_T$) and multiple filters ($N_F$) respectively. Together these parameters define the model transit lightcurve.

There are many systematic trends which might be introduced over the course of observing that cannot be accounted for using the reduction protocol described in $\S$ \ref{sec-imred}. For example, differential extinction due to airmass variation or photometric variation due to centroids wandering over pixels of varying sensitivities on an imperfectly flatfielded image. So for each image we extracted a set of nuisance parameters which were then used to compute a correction function (\ie\ detrending function) to remove systematic trends. We modeled this function as a linear combination of nuisance parameters:
\begin{equation}
F_{cor,i} = a_{0} + \sum_{k=1}^{N_{nus}} a_{k} X_{k,i},
\label{eqn-det}
\end{equation}
where $X_{k,i}$ are the nuisance parameters and $a_{k,i}$ are the corresponding coefficients. A typical set of nuisance parameters included (i) the airmass, (ii) the centroid positions of the target and reference stars, (iii) the local sky around the target and comparison stars, (iv) the global sky and (v) the total counts in the area that defined the photometric aperture on the masterdark and (vi) the master skyflat.

For each night, the entire lightcurve is normalized to 1 by an initial best-fit transit model. Then we fit for the coefficients $a_{k}$ of the correction functions using these residuals and a generalized linear least squares minimizer. The $a_{0}$ term is set to a constant, which in this case is 1 because of the way the lightcurve is normalized. We found that the most significant systematic trends in the residuals were correlated with airmass and the counts on the masterflat. This tells us that differential extinction and imperfect flatfielding are the two greatest sources of systematic effects.

Once the correction function and the model are derived we compute the goodness of fit to our data as:
\begin{equation}
\chi^2 = \sum_{i}^{N_{Data}} \frac{(O_i - M_i(\boldsymbol{\theta}) - F_{cor,i})^2}{\sigma_i^2},
\label{eq-chi2}
\end{equation}
where the $O_i$ and $\sigma_i$ are the observed data and associated errors, $M_i$ is the transit model and $F_{cor,i}$ is the detrending function. For all subsequent optimization with either Markov Chains or a Non-Linear Minimizer, Equation \ref{eq-chi2} is used to evaluate the goodness of fit. One must note that we devised routines such that the correction function is recomputed along with the model lightcurve for each step in the Markov chain or each iteration in the Non-Linear Minimizer.

In $\S$ \ref{sec-stellaractivity} we note the possible evidence for stellar activity in some of our lightcurves. These features were seen in lightcurves after reductions and detrending, and we believe they are not associated with any systematic effects in the data. We excluded these points from our analysis of system parameters.

\subsection{Markov Chain Monte Carlo}
\label{sec-mcmc}
Bayesian inference techniques like Markov Chain Monte Carlo have become a popular tool for constraining system parameters from observational data. We used the Metropolis-Hastings (M-H) algorithm, a well known MCMC method, to constrain the uncertainties for fitted parameters in our transit model. \citet{tegmark04} and \citet{ford05} have very good descriptions of the algorithm and its application to relevant astronomical data. Our prescription is closest to that described by \citet{ford05}. We approximate the posterior distribution and joint probability distribution of our model parameters given the observed data \textbf{\textit{O}}, as $P(\boldsymbol\theta\mid\boldsymbol{O}) \propto P(\boldsymbol\theta)P(\boldsymbol{O}\mid\boldsymbol{\theta}) \propto e^{-\chi(\boldsymbol{\theta})^2/2}$. Our MCMC routines use the standard stepping and selection rules of the M-H algorithm. The jump functions for our parameters are of the form,
\begin{equation}
\boldsymbol{\theta}_{j+1} = \boldsymbol{\theta}_{j} +  G(0,\boldsymbol{\sigma}_\theta^2) f
\label{eq-step}
\end{equation}
where $\boldsymbol\theta$ and $\boldsymbol{\sigma}_{\theta}$ are the vectors of model parameters and their associated step-sizes respectively and $G(0,\sigma_\theta^2)$ is a random number drawn from a normal distribution with a mean of 0 and a variance of $\sigma^2_{\theta}$. The factor $f$ is an adaptive step-size controller which is used to guide the chain to the optimal acceptance rate. For the case where a jump is performed for the entire vector of model parameters, it has been shown that the optimal acceptance rate is $\sim$ 23\% \citep{gelman03}. This desired rate is achieved by adjusting the step-size controller ($f$) every 100 accepted steps according to $f_{new}$ = 434 $f_{old}/N_{trials}$, where $N_{trials}$ are the number of steps attempted for the last 100 accepted steps (see also \citet{colliercameron07}).

By varying the entire vector of model parameters and applying a single step-size modifier we risk the situation of using mismatched step-sizes and undersampling targeted posterior distributions. Statisticians have shown that well constructed chains will properly sample posterior distributions at the correct acceptance rate \citep{gelman03}. However, the acceptance rate is guided by the location of the chain in parameter space. The M-H algorithm's basis for sampling the posterior distribution lies in the fact that steps in low probability (large $\chi^2$) regions of parameter space are accepted less often than those in high probability regions. For example, if we make a poor choice and select a starting step-size for one parameter to be too large relative to the other parameters in the ensemble, this parameter will traverse between regions of high and low probability faster than the rest. The acceptance rate in this case is biased by jumps in one parameter, and the chain will have poorly sampled posterior distributions for the remaining parameters. The key to a well-constructed chain is to choose the relative starting stepsizes for parameters such that they \textit{all} roam high and low probability regions of parameter space at roughly the same rate. To find this ideal set of starting step-sizes we ran a set of exploratory chains (40,000 iterations), stepping only one parameter at a time, until the step-size controller settled the simulation to an acceptance rate of $\sim 44\%$ \citep[the optimal acceptance rate for the one-dimensional case,][]{gelman03}. Most of the chains reached this acceptance rate at an iteration between $\sim$1000-7000. The values of the adaptive stepsize controller ($f$) near the end of these short runs gave us an idea of how much we under/over  estimated the starting stepsize for a given parameter. Typically, these ``final'' $f$ values multiplied by the starting stepsize (for the short runs) proved to be good choices for the starting stepsizes to be used for the larger, multi-parameter MCMC runs.

We ran 10 long MCMC chains for different combinations of parameter sets and lightcurves. Table \ref{table_mcmcstats} lists the names of the chains, the corresponding lightcurve data, the parameter sets used and some statistics from our post-run analysis of the chains. The chains are numbered from 001 to 005 which represent the 5 different data sets used -- listed in column 2. Chains 001, 004 and 005 are single-filter data sets corresponding to APOSTLE, MEarth and FLWO1.2m observations respectively. These were used to test the models where the transit depth $D$ was fit, for the highly limb-darkened case (see $\S$ \ref{sec-model-det}). Chain 002 was simply a redo of \citet{charbonneau09}'s analysis with our transit model and MCMC framework. In chain 003 we fit for transit parameters using all 3 data sets. The tags `a' and `b' denote whether the limb-darkening coefficients were left fixed or open respectively. For chains where the limb-darkening parameters were fixed (`a') we either chose values from the literature or used values which were found to be suitable by others. For the APOSTLE \rfilter\ dataset we chose values from \citet{claret04} for a 3000K star: ($v_{1,APOSTLE},v_{2,APOSTLE}$) = (0.908,0.305). For the MEarth and FLWO1.2m data we used unpublished values used by \citet{charbonneau09} in their fit (priv. comm. P. Nutzman). For the MEarth ``filter'' we used ($v_{1,MEarth},v_{2,MEarth}$) = (0.145,0.639) and for the FLWO1.2 z-band filter we used ($v_{1,FLWO1.2m},v_{2,FLWO1.2m}$) = (0.404,-0.289). Since it is well known that these parameters are highly degenerate, the `b' chains were run as a test of how well these lightcurves could be used to constrain stellar limb-darkening. The parameter set corresponding to each chain is listed in column 3 of Table \ref{table_pars1}. For the chains with multi-wavelength data, like 002 and 003, the parameter set $\theta_{2}$ was used, and for single filter data we used the set $\theta_{1}$ (see $\S$ \ref{sec-model-det}).

The typical number of iterations used for long MCMC chains ranged between 1-2.5$\times$10$^6$. These computations took a total of 80 CPU hours to complete on Linux workstations. For each chain, we selected only those steps where the acceptance rate remained roughly within 5\% of the optimal acceptance rate. Cropping the chains in this manner meant, many of the initial steps were discarded as most chains started with acceptance rates which were far from the optimal rate.

The M-H algorithm's jump function is only dependent on the previous location of a step in the chain, so it is not unusual for sequential points in the chain to be correlated. The dimensionless autocorrelation function provides a good assessment of how many independent points there are in a chain. We computed this for each open parameter per chain, using the prescription of \citet{tegmark04}, and derived the correlation and effective lengths for each chain. These are presented in Table \ref{table_mcmcstats}. The effective length (the chain length divided by the correlation length) is a measure of the number of independent points in a chain and must be large ($\gg 1$) for the errors derived from a chain to be meaningful. A large number of independent points signifies a statistically significant, well-sampled posterior distribution. We found that chains where the limb-darkening was fixed (`a' chains) had the greatest effective lengths and chains where limb-darkening parameters were left open (`b' chains) had the shortest effective lengths (see Table \ref{table_mcmcstats}). Chain 003b had the lowest effective length. This chain also had the largest number of open parameters (including 3 pairs of limb-darkening parameters). We believe the large number of limb-darkening parameters resulted in slow convergence for this chain. We discuss the significance of the resulting parameters and errors in the following section.

\subsection{Parameters and Errors}
\label{sec-params}
System parameters derived from our analysis are presented in Table \ref{table_pars1} and Table \ref{table_pars2} for the single and multi-wavelength datasets respectively. Directly fit model parameters are listed in the table sub-section titled `Model'. These correspond to the variables described in section $\S$ \ref{sec-model-det} as part of the parameter sets $\boldsymbol{\theta_{1}}$ and $\boldsymbol{\theta_{2}}$. The values listed in the tables were obtained using the minimization package MINUIT \citep{james94}. The $\chi^2$ from the best-fit model and the degrees-of-freedom (DOF) are listed in the last two columns of Table \ref{table_mcmcstats}. The ensemble of points from the posterior distributions of `Model' points were then used to compute posterior distributions of various `Derived' parameters \citep{seagermalleno03,carter08}. We chose to present the following 7 derived quantities: \textbf{i.} the planet-to-star radius ratio ($R_{p}/R_{\star}$), \textbf{ii.} the orbital period, \textbf{iii.} the impact parameter ($b$), \textbf{iv.} semi-major axis in stellar radius units ($a/R_{\star}$), \textbf{v.} orbital inclination ($i$), \textbf{vi.} orbital velocity ($v$) normalized by stellar radius, and \textbf{vii.} the stellar density ($\rho_{\star}$). The errors for both the `Model' and `Derived' parameters were then computed by sorting these data and choosing the 68.3\% confidence intervals to represent the 1$\sigma$ uncertainties. For cases where the parameter values were distributed asymmetrically around the best-fit value (at a level $>10\%$), we present upper and lower uncertainty estimates.

\begin{figure}
\centering
\epsfig{file=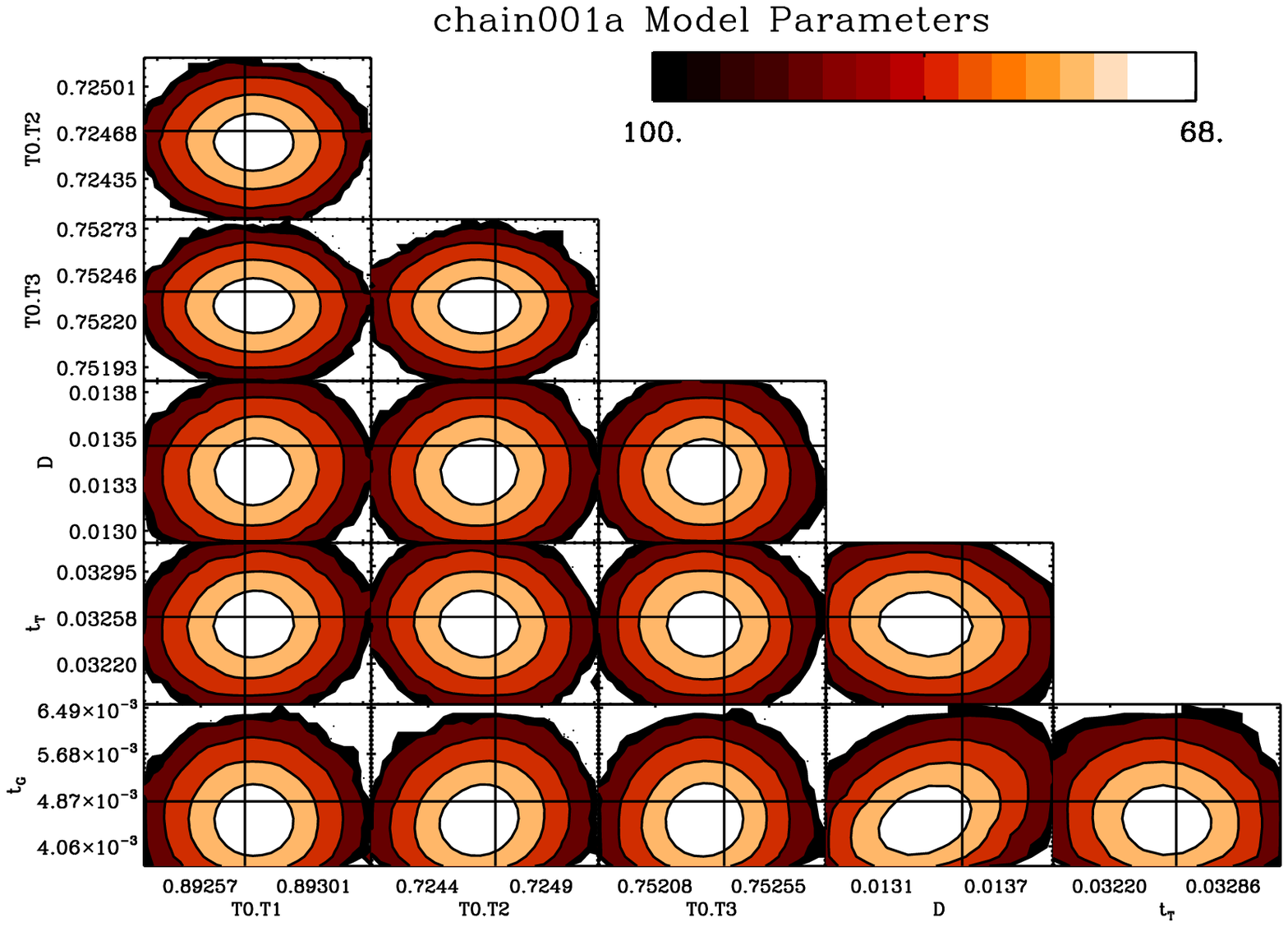}
\caption{\label{figure_mcmc_1} Joint probability distributions for all fitted transit `Model' parameters fit in MCMC chain001a. The model parameters were fit to APOSTLE lightcurves using the parameter set $\boldsymbol{\theta_{1}}$. The numbers and units correspond those listed in Table \ref{table_pars1}. The solid-line crosshairs mark the location of the best-fit values (also from Table \ref{table_pars1}). The contours mark the 1, 2, 3, 4 and 5 sigma regions, each enclosing 68.27$\%$, 95.45$\%$, 99.73$\%$, 99.994$\%$ and 99.99994$\%$ of the points in the distributions respectively.}
\end{figure}

\begin{figure}
\centering
\epsfig{file=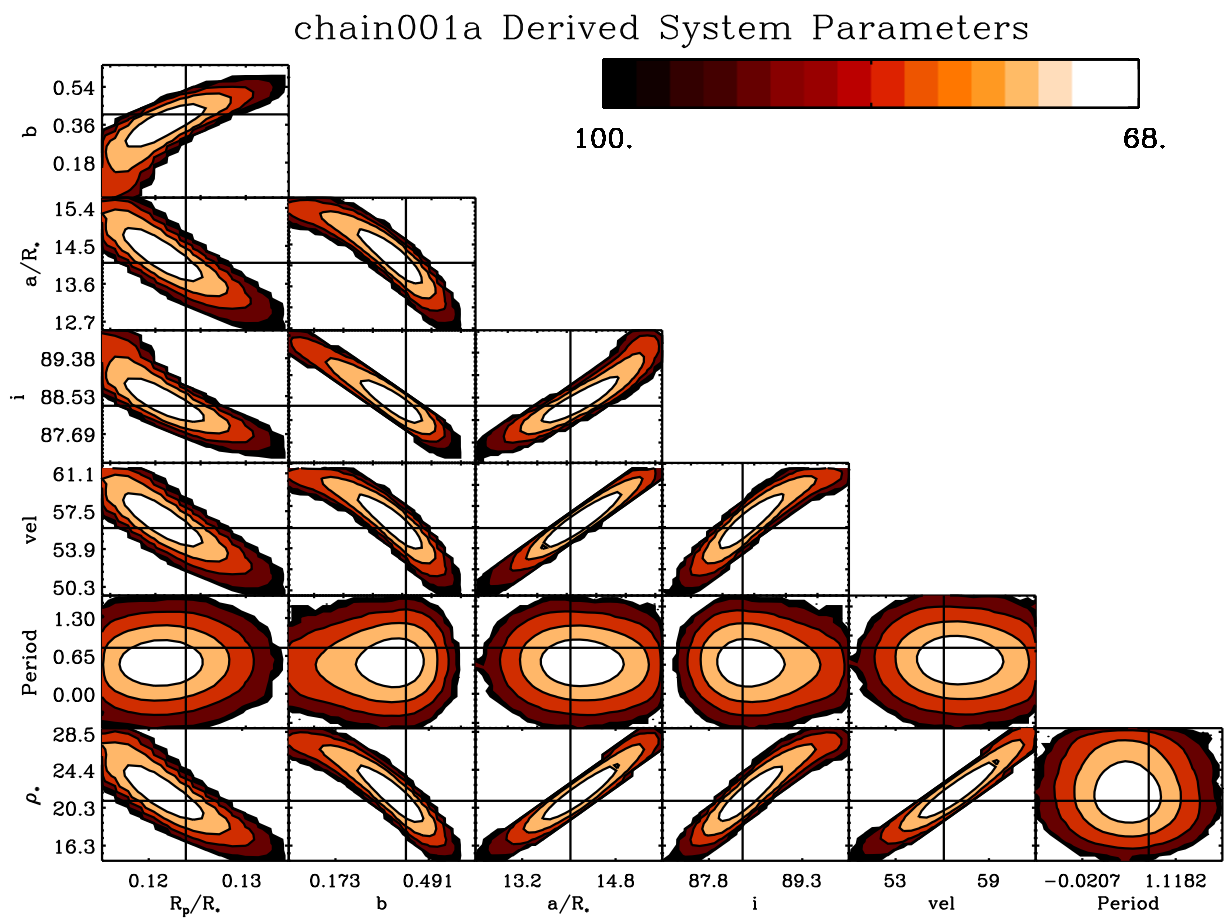}
\caption{\label{figure_mcmc_2} Joint probability distributions for all `Derived' parameters fit in MCMC chain001a. The numbers and units correspond those listed in Table \ref{table_pars1}. The solid-line crosshairs mark the location of the best-fit values (also from Table \ref{table_pars1}). The contours mark the 1, 2, 3, 4 and 5 sigma regions, each enclosing 68.27$\%$, 95.45$\%$, 99.73$\%$, 99.994$\%$ and 99.99994$\%$ of the points in the distributions respectively. Several parameters have strong mutual correlations.}
\end{figure}

Figure \ref{figure_mcmc_1} shows the joint-probability distributions of directly fit model parameters from chain001a. The cross-hairs on each sub-plot mark the best-fit values. We can see that the distributions of `Model' parameters shown in Figure \ref{figure_mcmc_1} show little or no correlation between each other. The use of parameters like transit duration ($t_T$) and ingress/egress duration ($t_G$) in our lightcurve model has resulted in a parameter set with few degeneracies. Figure \ref{figure_mcmc_2} shows results from the same chain, but of derived parameters. Here we can see that most of the derived parameters show strong correlations with each other. When characterizing lightcurves, it is fairly common practice to use some of the parameters listed here as `Derived' as the directly fit model parameters \citep[e.g.][]{holman06,winn07,colliercameron07}. Also common is the use of MCMC to obtain error estimates from such models. The presence of such correlations makes the interpretation of MCMC results challenging. These degeneracies can result in 1) chains that have short effective lengths (slow convergence) or 2) incomplete sampling of posterior probability distributions. Both problems can be reduced by running longer chains. However this solution is not always practical as the number of required steps may be prohibitively large or unknown. In $\S$ \ref{sec-mcmc} we noted the problem of slow convergence for chains when the limb-darkening was allowed to vary. Similarly, the problem of insufficient sampling of posterior distributions is common when there are ``banana-shaped'' degeneracies between parameters. In Figure \ref{figure_mcmc_2} we see such a degeneracy for the case of $b$ vs. $\rho_{\star}$, $b$ vs. $a/R_{\star}$ and $b$ vs. $R_{p}/R_{\star}$. In such situations a chain may get stuck in ``valleys'' of low $\chi^2$ and fail to sample other regions of parameter space. We have shown how both these issues can be avoided by choosing a parameter set that is free of mutual degeneracies. Figure \ref{figure_mcmc_1} graphically confirms this and column--6 (Corr Length) in Table \ref{table_mcmcstats} quantitatively establishes that chains with parameters that show no mutual correlations (chains `a') converge quickly.

\begin{figure}
\centering
\epsfig{file=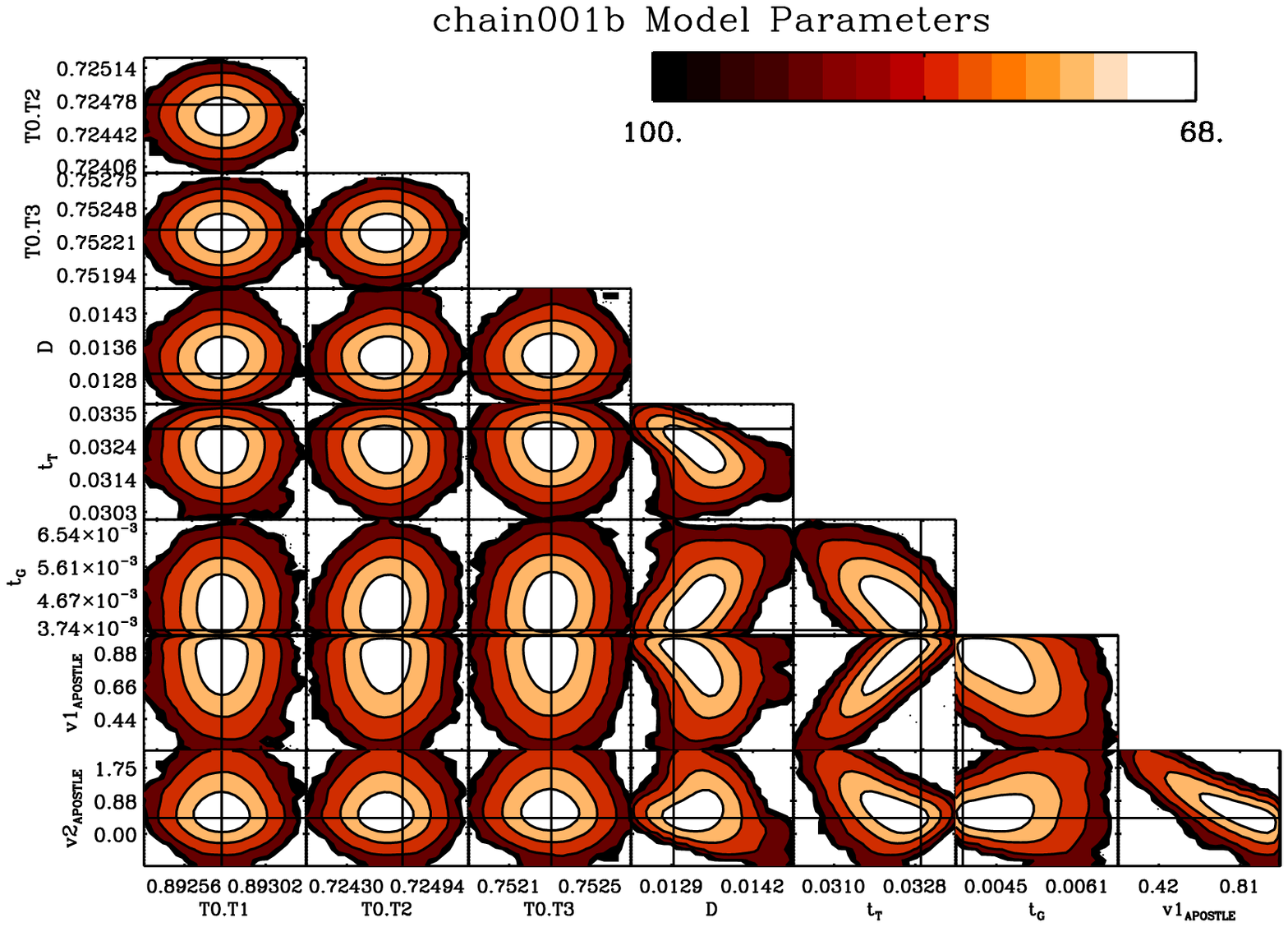}
\caption{\label{figure_mcmc_3} Joint probability distributions for all transit model parameters fit in MCMC chain001b. The model parameters were fit to APOSTLE lightcurves using the parameter set $\boldsymbol{\theta_{1}}$. The numbers and units correspond those listed in Table \ref{table_pars1}. The solid-line crosshairs mark the location of the best-fit values (also from Table \ref{table_pars1}). The contours mark the 1, 2, 3, 4 and 5 sigma regions, each enclosing 68.27$\%$, 95.45$\%$, 99.73$\%$, 99.994$\%$ and 99.99994$\%$ of the points in the distributions respectively. This set is different from chain001a, as the limb-darkening parameters $v1_{APOSTLE}$ and $v2_{APOSTLE}$ were added to the analysis. There are strong correlations seen between various model parameters (e.g. D vs $t_T$ and $t_T$ vs $v1_{APOSTLE}$). We can also see how the non-linear minimizer converges to a low value for $t_G$ and a high value for $t_T$. This seems to be a result of a degeneracy with the open $v1_{APOSTLE}$ parameter.}
\end{figure}

Figure \ref{figure_mcmc_3} shows our results from chain001b. This chain is similar to chain001a, except the limb-darkening parameters $v_1$ and $v_2$ were allowed to vary. As mentioned in $\S$ \ref{sec-model-det} we applied bounds ($v_1 + v_2 > 0$ and $0 < v_1 < 1$) to the limb-darkening parameters so as to avoid unphysical limb-darkening profiles. We note that correlations exist between the two limb-darkening parameters (Fig.\ref{figure_mcmc_3} panel -- $v_{1 APOSTLE}$ vs. $v_{2 APOSTLE}$). In addition, parameters that showed uncorrelated distributions in Figure \ref{figure_mcmc_1}, now show signs of being affected by degeneracies. The most strongly affected are the parameters $t_T$, $t_G$ and $D$. Not only do these parameters show degeneracies with the limb-darkening parameters, but they also show correlations between each other. The degeneracy between the ingress/egress duration ($t_G$) and the limb-darkening parameters can be understood simply by the fact that both affect the overall shape of the lightcurve. The limb-darkening can change the profile of the ingress and egress regions of the lightcurve, while $t_G$ determines the start and end points of ingress and egress. So given the error bars and scatter in our lightcurves, a range of limb-darkening and $t_G$ values can produce good fits. Our inability to constrain these parameters simultaneously seems to suggest that milli-magnitude photometry might not be sufficient for the complete characterization of transit lightcurves with limb-darkening. These degeneracies also strongly affect the best-fit parameters derived from these chains. For example, in the list of results for `b' chains in Tables \ref{table_pars1} and \ref{table_pars2}, we find that values for $t_G$ and $t_T$ produce very discrepant estimates for derived system parameters (like $i$, $b$ and $\rho_{\star}$). We conclude that results from the `b' chains are in general unreliable, so we elect to discuss results in which the limb-darkening parameters are held fixed.

\subsubsection{Which parameter set to consider?}
We analyzed a combination of five different datasets and two parameter options (chains `a' and `b', see Table \ref{table_mcmcstats}). Out of the five `a' chains three chains, 001a, 004a and 005a were single filter datasets (see Table \ref{table_pars1}). These serve as simple checks on our transit model and MCMC framework. They also highlight the use of the maximum transit depth ($D$) instead of the square of planet-to-star radius ratio ($R^2_{p}/R^2_{\star}$) for single filter data. We find that $D$ has weaker covariance with $t_{G}$ and $t_{T}$ than does $R^{2}_{p}/R^{2}_{\star}$. Due to the small number of lightcurves for each of these three sets we see, not surprisingly, that parameters such as the planet-to-star radius ratio ($R_{p}/R_{\star}$) and period are constrained to low precision. The chains numbered 002 were simply a redo of \citet{charbonneau09}'s analysis with our transit model and MCMC framework. We find the results are in good agreement with \citet{charbonneau09}'s findings. The largest parameter set presented in this work is chain003a. This set represents the joint analysis of multi-wavelength data using our transit model framework, and is in good agreement with the results presented in \citet{charbonneau09}. From Table \ref{table_mcmcstats} we can safely say that the MCMC analysis for this chain is quite reliable. \textit{Thus the results from chain003a are the best to consider from this work and are used in the discussions that follow.}

\section{Transit Timing Variations}
\label{sec-ttv}
\begin{figure}
\centering
\epsfig{file=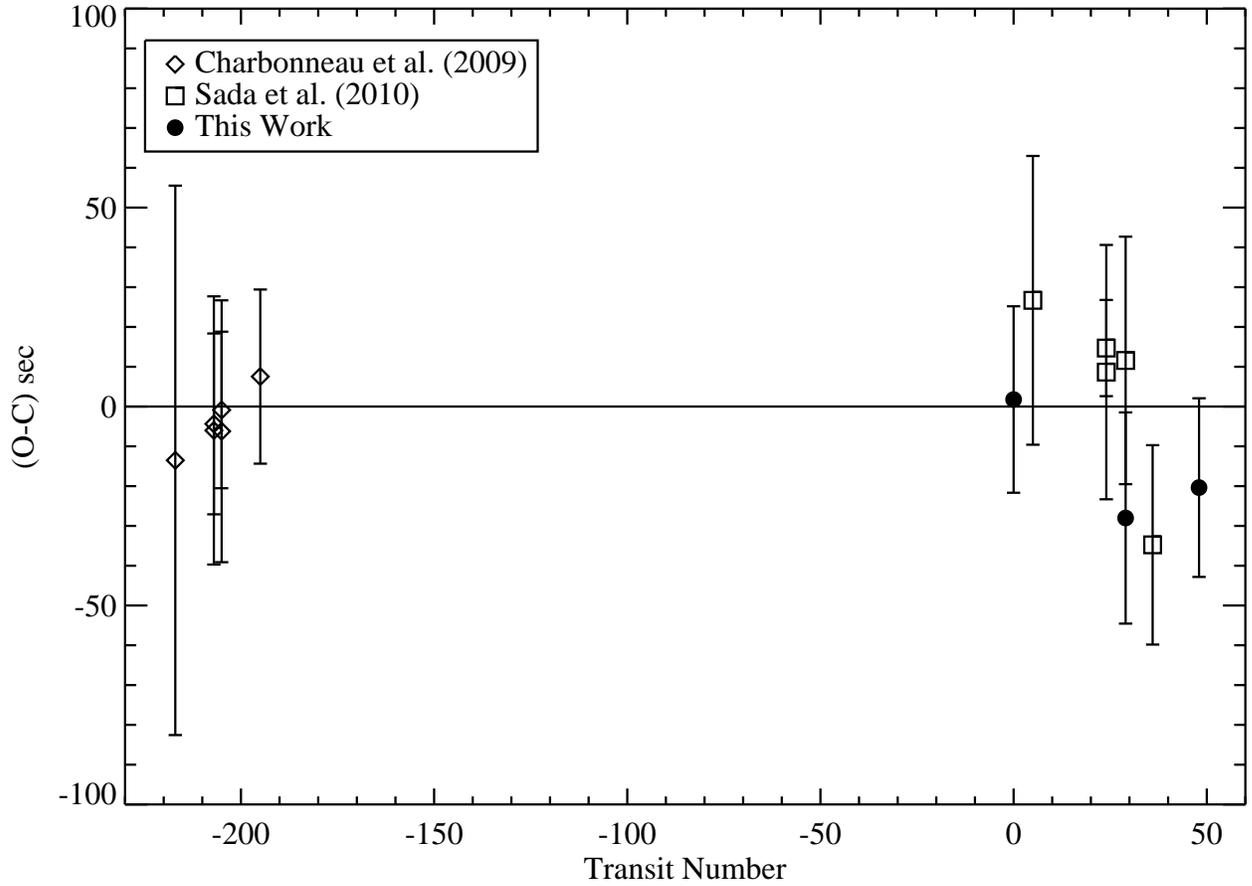}
\caption{\label{figure_ttv} The observed minus calculated (O-C) transit times versus transit number for GJ 1214b. All transit times were converted to TCB. The \citet{charbonneau09} times are from our analysis of the joint APOSTLE and MEarth dataset. The \citet{sada10} data were converted to BJD$_{TCB}$.}
\end{figure}

At the time of writing this paper, 10 transits have been reported for GJ 1214b \citep[including][]{charbonneau09,sada10}. Transit times reported in \citet{charbonneau09} and \citet{sada10} were in HJD$_{UTC}$, i.e. JD representation of UTC corrected for the light travel time delay to the solar system barycenter. These were converted to our preferred BJD$_{TCB}$ time coordinate. For the transits observed by \citet{charbonneau09}, in Table \ref{table_pars2} we present both the conversion of values cited by them and the transit times derived from our lightcurve fit to their data \citep[see columns `][' and `chain003a']{charbonneau09}. We chose to use the times which resulted from our analysis (chain003a) since we fit for many transits simultaneously. These BJD$_{TCB}$ times are not far off from the converted transit times. The largest discrepancy we found was $\sim15$ sec for the transit T0.T4. This difference is not surprising since it is the very first transit observed by \citet{charbonneau09} and so, has the least precise measurement ($\sigma \sim$40 sec). With all available transit times we refit the ephemeris for GJ 1214b:
\begin{equation}
\label{ephem}
TT(N_{Ti}) = T_{ZP}  + P \times N_{Ti}
\end{equation}
where TT is the expected transit time, T$_{ZP}$ is the zero-point of the transit times, $P$ is the orbital period and N$_{Ti}$ is the transit number. Given all the times and transit numbers, we fit for T$_{ZP}$ and P, and found them to be 2455307.892663$\pm$0.000082 BJD$_{TCB}$, and 1.58040487$\pm$0.00000067 days respectively. We set the zero-point to be the first transit observed by APOSTLE (T0.T1). In Figure \ref{figure_ttv} we show the observed minus calculated (O-C) transit times for these transits. The plot shows that there are no significant variations from the expected times of transit.

If this system did have additional planets they might induce variations in transit times \citep{agol05,holmanmurray05}. For the special case where these additional planets are in resonance with GJ 1214b, the resulting TTVs could be on the order of minutes and well above the precision limits of transit follow-up surveys. However such a signal does not seem to exist (Figure \ref{figure_ttv}). A null result however can be used to place interesting limits on the mass and orbital configurations of a possible undetected companion \citep{steffenagol05,agolsteffen07}. We will defer such an analysis until more data are available, as GJ 1214 remains part of APOSTLE's observing program.

\section{Absolute Stellar and Planetary Properties}
\label{sec-mr}
Transit and RV data when taken together allow us to constrain various planetary parameters. However, many of these directly depend on the estimate of absolute stellar parameters, the most important being absolute stellar mass and radius. Transit lightcurves allow us to constrain the average stellar density ($\rho_{\star}$). We can rewrite Kepler's 3rd law to get an expression of stellar density $\rho_{\star} = 3\pi (a/R_{\star})^3 / (G P^2)$, where $G$ is the universal gravitational constant \citep{seagermalleno03}. The parameters, $a/R_{\star}$ (the semi-major axis of the planet in stellar radius units), and $P$ (the orbital period) can be deduced from transit data \citep[see $\S$ \ref{sec-params},][]{carter08}. To translate this measurement into an estimate of absolute mass and radius an additional constraint is required. Usually one seeks an estimate of the stellar mass. Once the stellar mass is obtained, getting the stellar radius is trivial, since $\rho_{\star} = 3M_{\star}/(4\pi R^{3}_{\star})$.

Stellar mass is usually obtained using empirical or theoretical mass-luminosity relations of stars. However, given the stellar density measurement, one may also constrain mass and radius from the locus of points where the measured density (from transits) intersects with well-known stellar mass-radius relations. Mass-radius relations provide an understanding of the internal structures of stars. For stars in the mass range of GJ 1214, empirical mass-radius relations are difficult to interpret due to biases in the survey sample and large uncertainties in the measurement of stellar parameters. Theoretical considerations of internal structure show that the overall size of a star might be strongly affected by convection and magnetic activity \citep{chabrier07}. Mass-luminosity relations on the other hand stem from our understanding of the energy production in stars. Energy production rates from nuclear fusion in the core are better understood and hence the luminosity of low-mass stars relate better with mass \citep{CB97,hillenbrandwhite04}. In light of this, we describe the derivation of absolute stellar properties of GJ 1214 using mass-luminosity relations and discuss how the derived radius of GJ 1214 compares to well-known mass-radius relations of low-mass stars. Once the absolute stellar parameters are obtained, various measurements from RV and transit observations can be used to derive absolute parameters of GJ 1214b.

\subsection{Stellar Mass and Radius}
\label{sec-starmr}
\begin{figure}
\centering
\epsfig{file=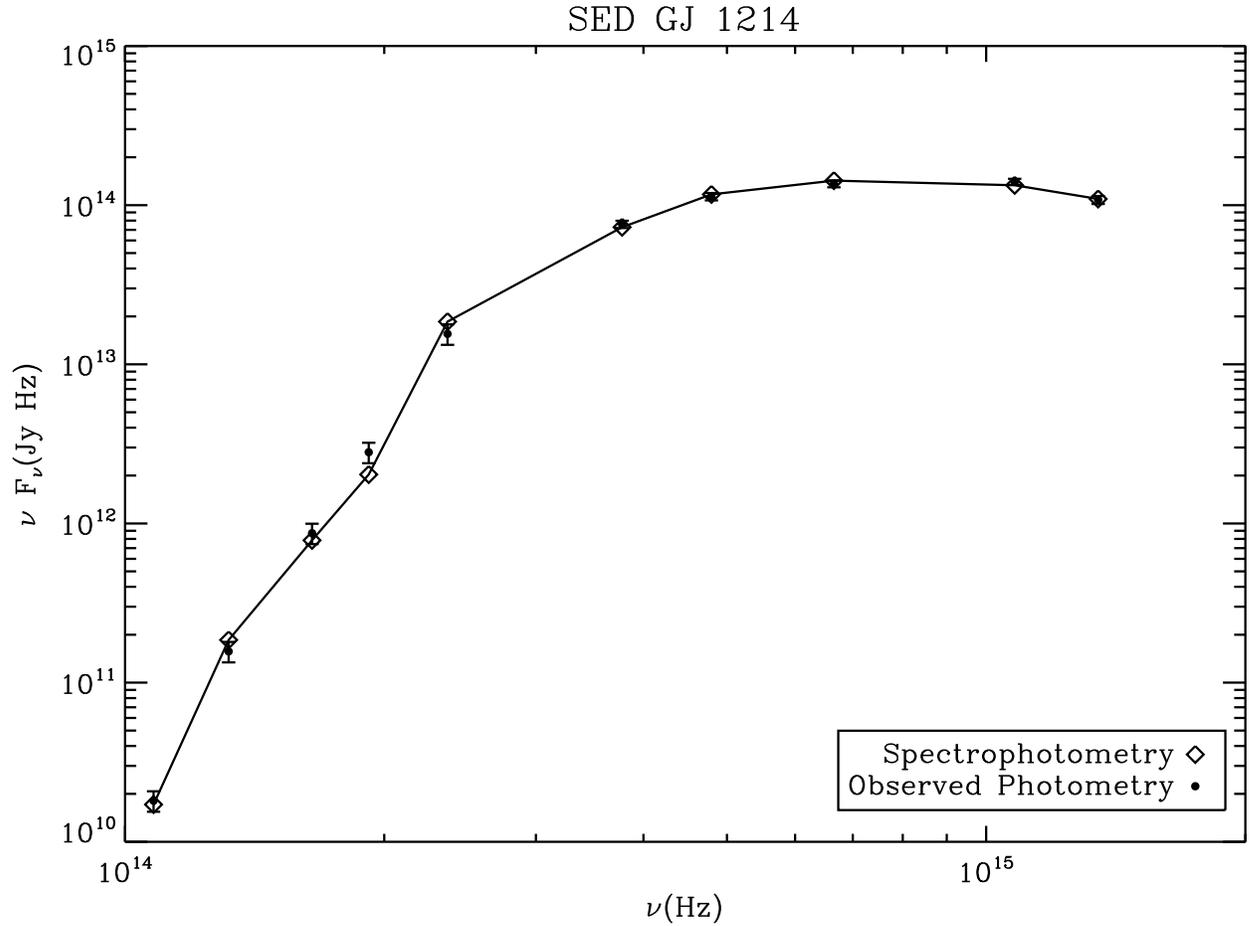}
\caption{\label{figure_sedplot} The spectral energy distribution (SED) of GJ 1214 and the resulting best-fit spectrophotometry as described in $\S$ \ref{sec-mr}. The photometric errors on the optical (UBVRI) data were adjusted to 15$\%$, while those for the infrared data (2MASS and Warm Spitzer) were adjusted to 5$\%$.}
\end{figure}

\begin{figure}
\centering
\epsfig{file=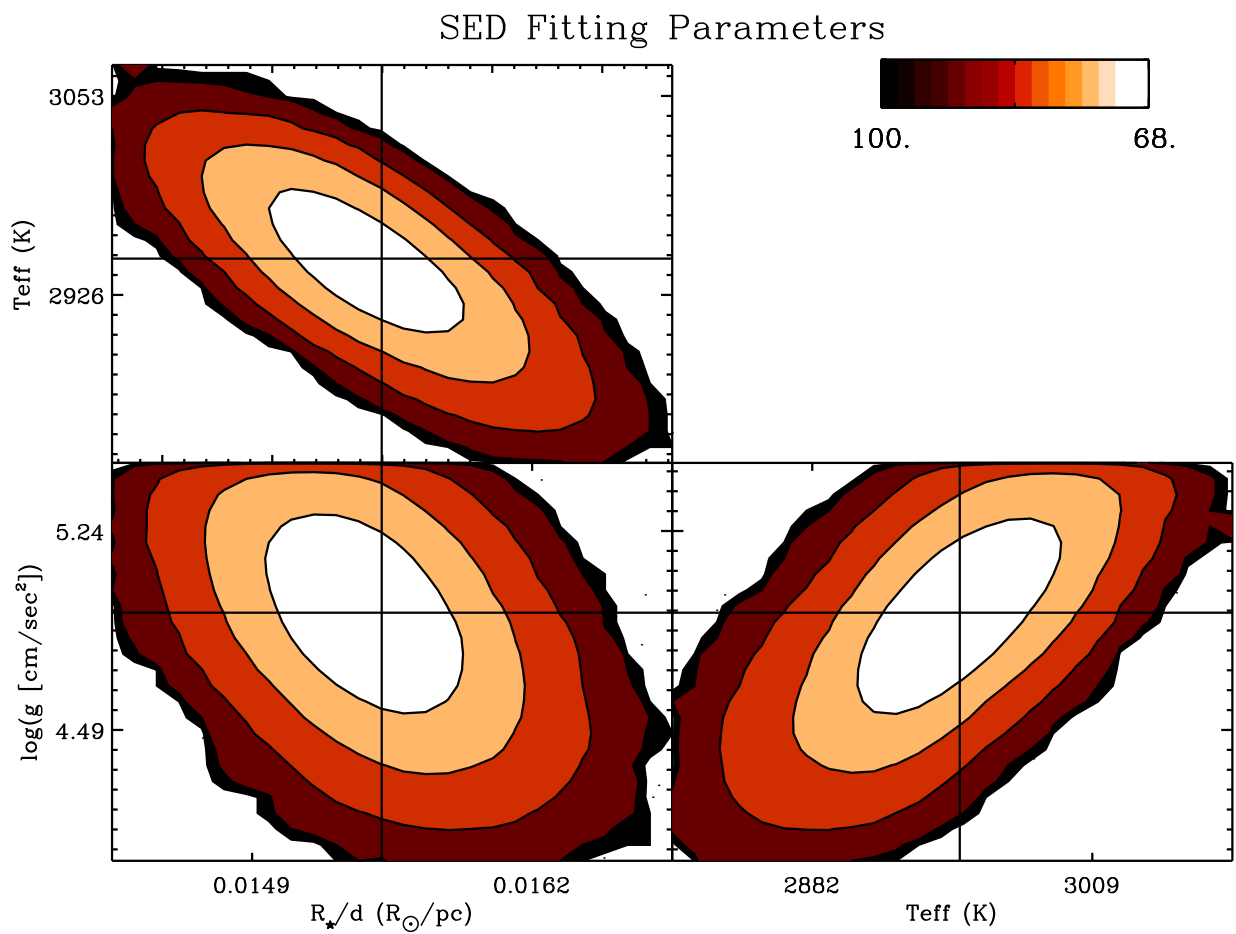}
\caption{\label{figure_mcmc_sed} Joint probability distributions for the three parameters used for SED fitting. The numbers and units correspond to those listed in Table \ref{table_pars3}. The solid-line crosshairs mark the location of the best-fit values (also from Table \ref{table_pars3}). The contours mark the 1, 2, 3, 4 and 5 sigma regions, each enclosing 68.27$\%$, 95.45$\%$, 99.73$\%$, 99.994$\%$ and 99.99994$\%$ of the points in the distributions respectively.}
\end{figure}

Photometry is available in 8 wavebands (UBVRIJHK) for GJ 1214 \citep{DF92,cutri03}. We also obtained unpublished IRAC1 and IRAC2 flux estimates from Jean-Michel D\'{e}sert (priv. comm.). Together these data cover the peak of GJ 1214's spectral energy distribution (SED). This allowed us to fit the observed SED with spectrophotometry \citep[derived using the technique described in][]{ivezic07,maiz06} from model spectra \citep{hauschildt99}. The errors on the optical photometry were increased to 15$\%$ and the infrared data to 5$\%$ to compensate for inaccuracies in spectrophotometry. We fit for the effective temperatures (T$_{eff}$) and $\log{g}$ over the model grid. In addition we fit for a constant, $R_{\star}/d$, which is the ratio of the stellar radius to the distance of the star from Earth. This factor is related to the solid angle of the star and scales the synthetic spectra to the observed fluxes. Our best fit produced a $\chi^2$ of 10.5 given 7 degrees-of-freedoms (see Figure \ref{figure_sedplot}). The uncertainties on the fit parameters were derived using the MCMC technique described in $\S$ \ref{sec-mcmc}. Results from the MCMC run are shown in Figure \ref{figure_mcmc_sed}. We integrated the resulting best-fit spectra (extracted at the best-fit $T_{eff}$ and $\log{g}$ on the grid) over all wavelengths and scaled it with the solid-angle (related to $R_{\star}/d$) to derive the observed flux ($F_{obs}$). We then used the parallax quoted by \citet{vanAltena95} to get an estimate of GJ 1214's bolometric luminosity, $L_{\star} = 0.0028\pm0.0004 L_{\sun}$. Parameter values and uncertainties for all fit and derived parameters are quoted in Table \ref{table_pars3}.

\begin{figure}
\centering
\epsfig{file=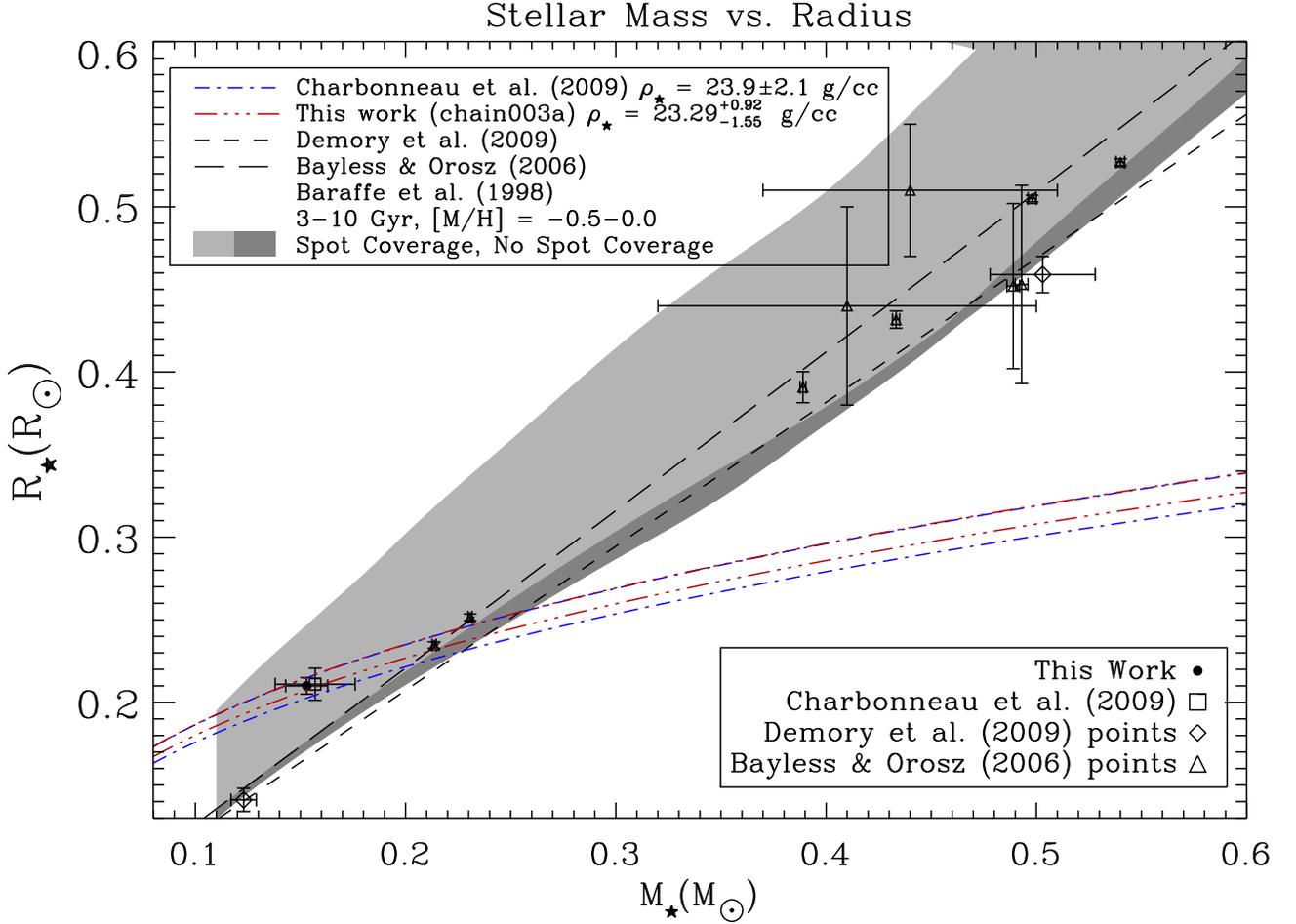}
\caption{\label{figure_mrplot} Stellar mass (M$_{\star}$) vs. stellar radius (R$_{\star}$), in solar units. The linestyles are matched to their corresponding references in the legend on the top-left corner of the figure and the legend on the bottom right matches the data points on the plot. The two \textit{dashed lines} in \textit{black} are empirically derived mass-radius relations for low-mass main-sequence stars \citep{demory09,baylessorosz06}. The data points are various estimates of mass and radius. The estimate of mass and radius presented in this work and \citet{charbonneau09} are also marked on the plot. The two \textit{dashed-dotted} curves, shown in \textit{red} and \textit{blue} in the color version, are the 1$\sigma$ contours of constant stellar density obtained from transit observations in this work and \citet{charbonneau09} respectively. The shaded regions represent the spread in stellar mass and radius taking into account various theoretical considerations. The \textit{darker} region represents the spread over age (3--10 Gyr) and metallicity (-0.5--0.0) for stars without spots. The \textit{lighter} region shows this spread when spot-coverage is introduced using the formalism of \citet{chabrier07}. The outer limit (leftmost edge) of the \textit{light} region represents the extreme case of 100$\%$ area coverage of spots which are 500K cooler than the fiducial the surface temperature of the star for a given mass \citep{morales10}.}
\end{figure}

With an estimate of $L_{\star}$, theoretical or empirical mass-luminosity relations can be used to estimate the stellar mass. We used data presented by \citet{baraffe98} for solar metallicity stars at an age of 5 Gyr and found GJ 1214 to have a mass $M_{\star} = 0.153 \pm 0.010 M_{\sun}$. GJ 1214's age is assumed to lie somewhere between 3--10 Gyr since its measured kinematics place it in the old disk population \citep{reid95}. The spread in mass-luminosity over this age range is insignificant ($\ll 1\%$) so mass and luminosity data from the 5 Gyr isochrone served very well. GJ 1214's metallicity has not been directly measured, but indirect means indicate that [Fe/H] $>$ 0.0. For example, photometric proxies indicate that its [Fe/H] = +0.03 or +0.28 \citep[][respectively]{johnsonapps09,schlaufman10}. \citet{rojas-ayala10} use near-IR equivalent widths of NAI, CaI and the H$_{2}$O index \citep{covey10} for M-Dwarfs with and without planets to derive a spectroscopic metallicity indicator for M-Dwarfs. Using this method they report an [Fe/H] = +0.39. The \citet{baraffe98} models unfortunately do not cover this high metallicity range, so as a check we also estimate the stellar mass using an empirical mass-luminosity function presented by \citet{scalo07}. This fit is based on masses derived from observations of binary stars \citep{hillenbrandwhite04}, and the sample should represent a wide range of metallicities. Using this fit we derived $M_{\star} = 0.148 M_{\sun}$, which is $0.005 M_{\sun}$ lower than the value derived from the theoretical relation but well within the uncertainties. \citet{charbonneau09} use a similar technique but use a mass-luminosity function calibrated for mid-infrared K-band luminosities to determine the mass \citep{delfosse00}. Their estimate of the mass was 0.157 $\pm$ 0.019 M$_{\sun}$ and is consistent with our measurement. Estimating the stellar radius once the mass is estimated is trivial and Figure \ref{figure_mrplot} demonstrates how this is done. The two \textit{dashed-dotted} curves on the plot of stellar mass vs. radius (shown in \textit{red} and \textit{blue} in the color version), are the 1$\sigma$ contours of constant stellar density obtained from transit observations in this work and \citet{charbonneau09} respectively. The point where the mass estimate intersects the measured density gives us the stellar radius, and the points of intersection with the 1$\sigma$ density contours determine the errors on absolute stellar mass and radius (see Table \ref{table_pars3}).

Figure \ref{figure_mrplot} shows that our estimates for the mass and radius of GJ 1214 do not lie close to the empirical mass-radius relations for low-mass stars \citep{demory09,baylessorosz06}; the empirical relations are represented by the \textit{dashed lines} in the figure. There is also disagreement between the derived values and the theoretical relations when there is no spot coverage. The \textit{dark shaded region} on the figure shows the spread in mass and radius due to a range of metallicities and ages from theoretical models \citep{baraffe98} and no star spots. Age and metallicity alone cannot account for the discrepent data. \citet{chabrier07} discuss how the internal structure of low-mass stars are affected by magnetic activity and convection. For example, the overall sizes of stars might be affected by cool spots on the surface where magnetic field lines penetrate deep into the convective layer. This situation can reduce the efficiency of convective energy transport, causing the star to settle to a larger radius for a given luminosity and fiducial temperature \citep{morales10}. Using the formalism introduced by \citet{chabrier07} we computed the radii for the extreme case when the surface of a given star is assumed to be completely covered by spots which are cooler than the fiducial surface temperature by 500K. Figure \ref{figure_mrplot} shows the region bounded by this extreme as the \textit{light shaded region}. The variation in radius for a given stellar mass is a very strong function of spot-coverage. The derived absolute parameters fall within this region and suggest that GJ 1214 may have cool regions on its surface. GJ 1214 would have to have 92$\%$ of its surface covered by spots 300K cooler than its surface to explain the estimated radius. For spots cooler than the surface by 500K, the large radius may be explained by spots covering 61$\%$ of the stellar surface area. We discuss some evidence for spots and magnetic activity in a subsequent section ($\S$ \ref{sec-stellaractivity}). Addressing convection is beyond the scope of this paper.

An alternate explanation for the discrepancy could be the estimate of GJ 1214's luminosity. If empirical mass-radius relations are to be believed GJ 1214's mass and radius would be closer to 0.23 $M_{\sun}$ and 0.24 $R_{\sun}$, based on the intersection with the density contours. Working backwards, this translates to a bolometric luminosity of roughly 0.0065 $L_{\sun}$. This is a very large difference in luminosity from the current estimate. The greater luminosity can only be reconciled with the observed fluxes if it was further away by 7pc (i.e. a parallax that was smaller by 0.027$\arcsec$). \citet{vanAltena95} measure the parallax for GJ 1214 to be 0.0772$\pm$0.0054$\arcsec$. The discrepent measurement would have to be a 5$\sigma$ systematic error and hence highly unlikely.

\subsection{Absolute Planetary Parameters}
With the semi-amplitude ($K$) from RV measurements, and the Period ($P$) and inclination from transits observations, one can determine the planetary mass: $K \propto M_{p} \sin{i}/ ( M^{2/3}_{\star} P^{1/3})$. The planetary radius can be determined from the planet-to-star radius ratio measured from the transit lightcurve. From the planetary mass and radius, we derive the planet's density ($\rho_p$), escape velocity ($V_{esc,p}$) and surface gravity ($g_{p}$). Using the semi-major axis ($a/R_{\star}$), $R_{\star}$ and stellar effective temperature ($T_{eff}$) we estimated the equilibrium temperature of the planet assuming a Bond albedo of 0.0 and 0.75. All errors were propagated assuming Gaussian uncertainties. We list our estimates in Table \ref{table_pars3}. The planetary mass, radius and density we derive are $M_{p} = 6.37\pm0.87 M_{\earth}$, $R_{p}= 2.74_{-0.05}^{+0.06} R_{\earth}$, and $\rho_{p} = 1.68\pm0.23 g/cm^3$, respectively. These data confirm GJ 1214b's status as a Super-Earth and the density measurement attests the presence of a massive gas envelope.

\section{Stellar Activity}
\label{sec-stellaractivity}
Many main-sequence stars are believed to be magnetically active and the frequency of active stars is known to increase with decreasing mass. Most observed variability on such stars is likely due to star spots and stellar flares \citep{basri10,walkowicz10}. The evidence for spots on the stellar surface is inferred from rotationally modulated long--term periodic trends in stellar lightcurves. Flares on the other hand are short--term events which are believed to be caused by the sudden release of energy from the reconnection of magnetic field lines near an active surface region. Some of the most active M-Dwarfs are believed to lie at masses below the transition between partially and fully convective interiors \citep[$< 0.35 M_{\sun}$,][]{reinersbasri09}. The increased activity is often thought to be a result of asymmetric magnetic field topologies for fully convective low mass stars. Although GJ 1214 (0.153$M_{\sun}$) lies well within this mass range, it has been classified as an inactive M-Dwarf by \citet{hawley96} based on a H$\alpha$ activity index. Active stars are bound to have numerous cool or hot regions on their surface; the existence of such regions can be inferred from the detection of spot or flare features in lightcurves. As we have already discussed in $\S$ \ref{sec-mr} the absolute size of GJ 1214 is not in agreement with our understanding of the radii of spot-free stars.

\begin{figure}
\centering
\epsfig{file=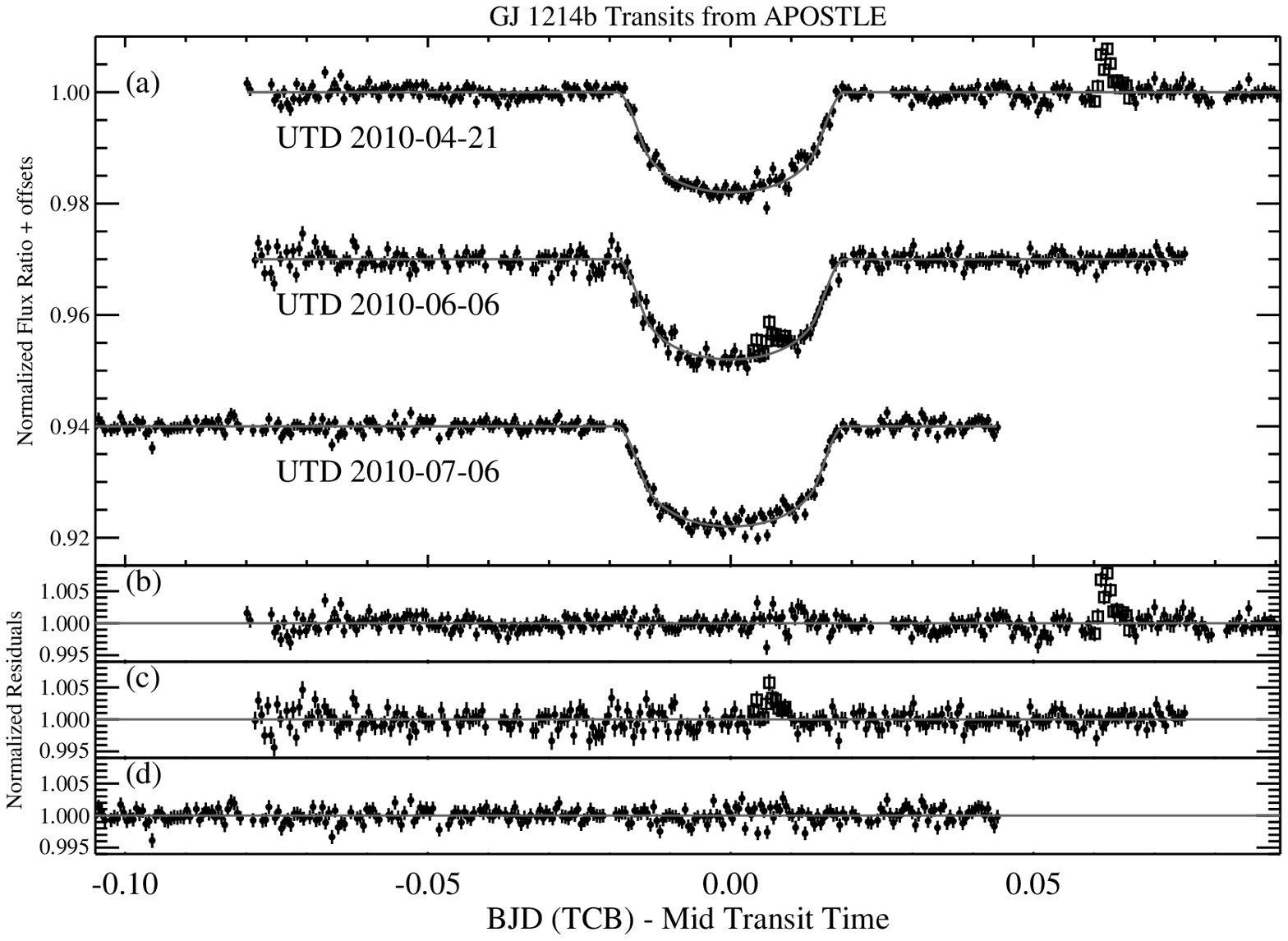}
\caption{\label{figure_lcs} (a) Lightcurves of 3 transits of GJ 1214b observed by APOSTLE. The vertical axis shows the normalized flux ratio and the horizontal axis shows time from mid-transit time in days. The transit time (T0) estimated for each transit was subtracted from the time stamp for each point. The solid circles represent the points used in the analysis, while the open squares represent the data that were excluded. The excluded data were either part of the stellar flare (UTD 2010-04-21) or the possible spot-crossing event (UTD 2010-06-06). The gray line shows the transit model. Panels (b), (c) and (d) show the residuals from the lightcurves normalized by the transit model. The typical scatter in the normalized flux ratio (excluding the flare, spot and transit signal) was $\sim 0.0011$.}
\end{figure}

\begin{figure}
\centering
\epsfig{file=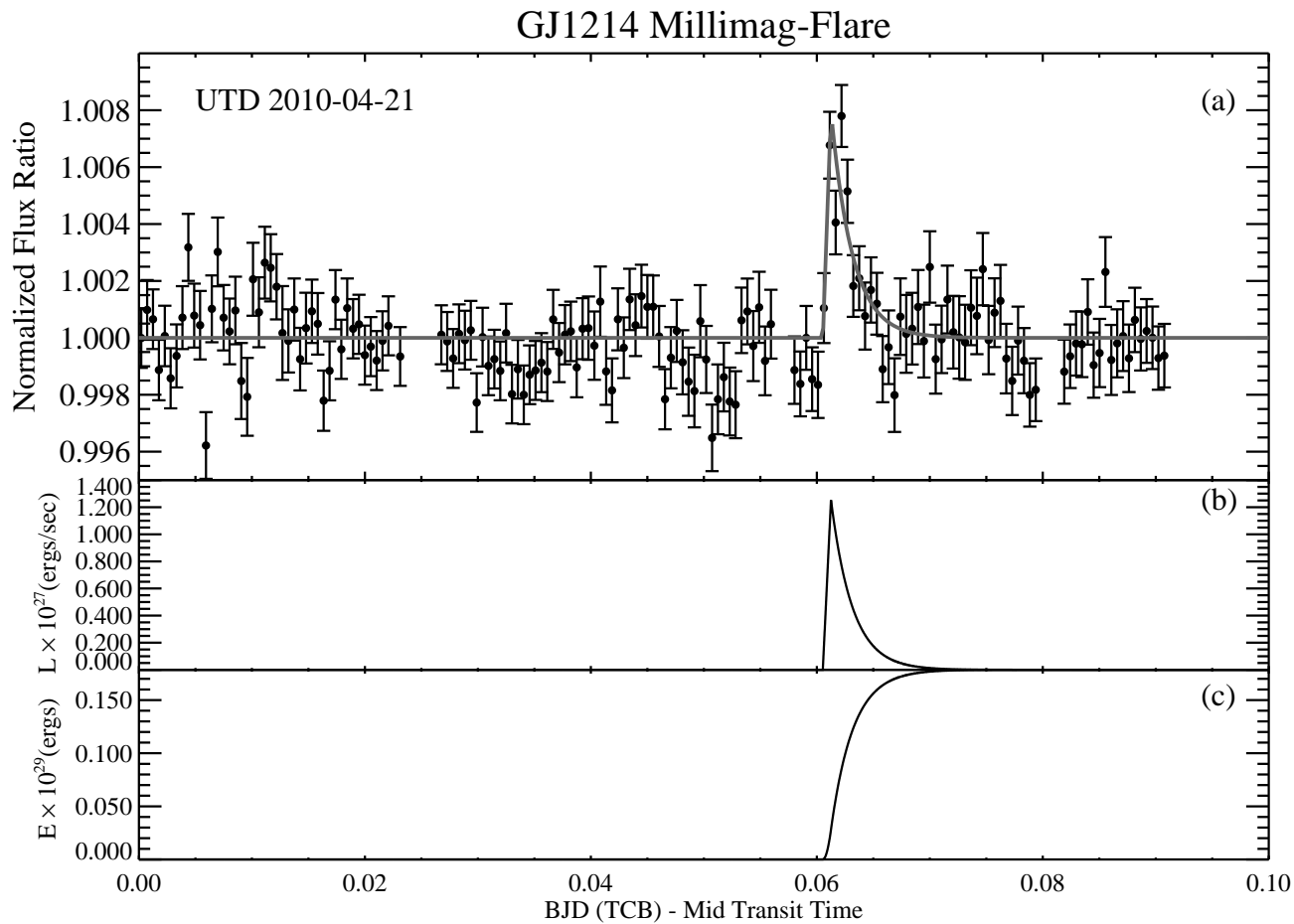}
\caption{\label{figure_flare} (a) Shows the flare event observed on UTD 2010-04-21. The data are the same as those shown in Figure \ref{figure_lcs} (b), only magnified. The gray line is the best-fit FRED model. Even though only the points after mid-transit are shown, the fit was made using all lightcurve points (in Figure \ref{figure_lcs} (b)). Panel (b) in the above figure shows the \rfilter\ luminosity as a function of time and panel (c) shows the total energy output by the flare above the quiescent \rfilter\ level as a function of time.}
\end{figure}

We believe that our \rfilter\ observations show evidence for a low-energy stellar flare on GJ 1214 and the possible detection of a cool spot on the surface. The sharp rising and falling trend seen in the out-of-eclipse lightcurve on UTD 2010-04-21 (Figure \ref{figure_lcs}) is similar to the fast-rise exponential decay (FRED) shape commonly associated with stellar flares \citep{hawleypettersen91}. Panel (a) in Figure \ref{figure_flare} shows this event in greater detail. We built a lightcurve model with two components: 1) a linear rise phase and 2) an exponential decay phase. We fit for the start time, peak time, peak flux and the e-folding time of the exponential phase. The best-fit lightcurve is shown as the solid gray line in Figure \ref{figure_flare} (a). The $\Delta\chi^2$ for this flare model compared to a straight line fit to the data is 112.6. The \rfilter\ flux of the star rose to a peak 0.8\% above the quiescent level and decayed over $\sim$3 minutes (e-folding time). Since we lacked flux-calibrated photometry, we used synthetic stellar spectra \citep{hauschildt99} to estimate the energy output by this event. We determined the \rfilter\ flux by integrating the synthetic spectra of a star with $T_{eff} = 2949K$ and $\log{g} = 4.94$ over the spectral response of the \rfilter\ \citep{ivezic07,maiz06}. Using a stellar radius of 0.21$R_{\sun}$, we computed the quiescent luminosity in the \rfilter\ to be $\sim$ 1.6$\times$10$^{29}$ ergs/sec. Panel (b) shows our flare model in luminosity units above the quiescent level for GJ 1214. Following the method described in \citet{hawley03} and \citet{kowalski10} we integrated under the flare lightcurve and estimated the total energy output by the flare in the \rfilter\ to be $\sim$1.8$\times$10$^{28}$ ergs, see Figure \ref{figure_flare}(c). The time it would take for the non-flaring star to emit this amount of energy (refered to as the \textit{equivalent time} in the M dwarf flare community) in is 0.113 seconds. Compared to typical M-Dwarf flares, this event is short-lived and of much lower energy. In fact such events are likely to be drowned out by noise for most flare monitoring campaigns as milli-mag precision is not commonly desired when looking at the most active stars. \citet{hawley03} reported flare energies between 8--58 $\times$10$^{30}$ ergs from Johnson R-filter observations of the active star AD Leonis. The activity observed on GJ 1214 is 4 orders of magnitude lower in energy than some of the energetic flares observed on AD Leo (also an M4.5V star). AD Leo has been identified as a member of the young galactic disk population \citep{montes01}. \citet{west08} have established that stellar activity decreases with age, and hence the differences in the activity levels of GJ 1214 and AD Leo might be purely due the differences in their ages.

\begin{figure}
\centering
\epsfig{file=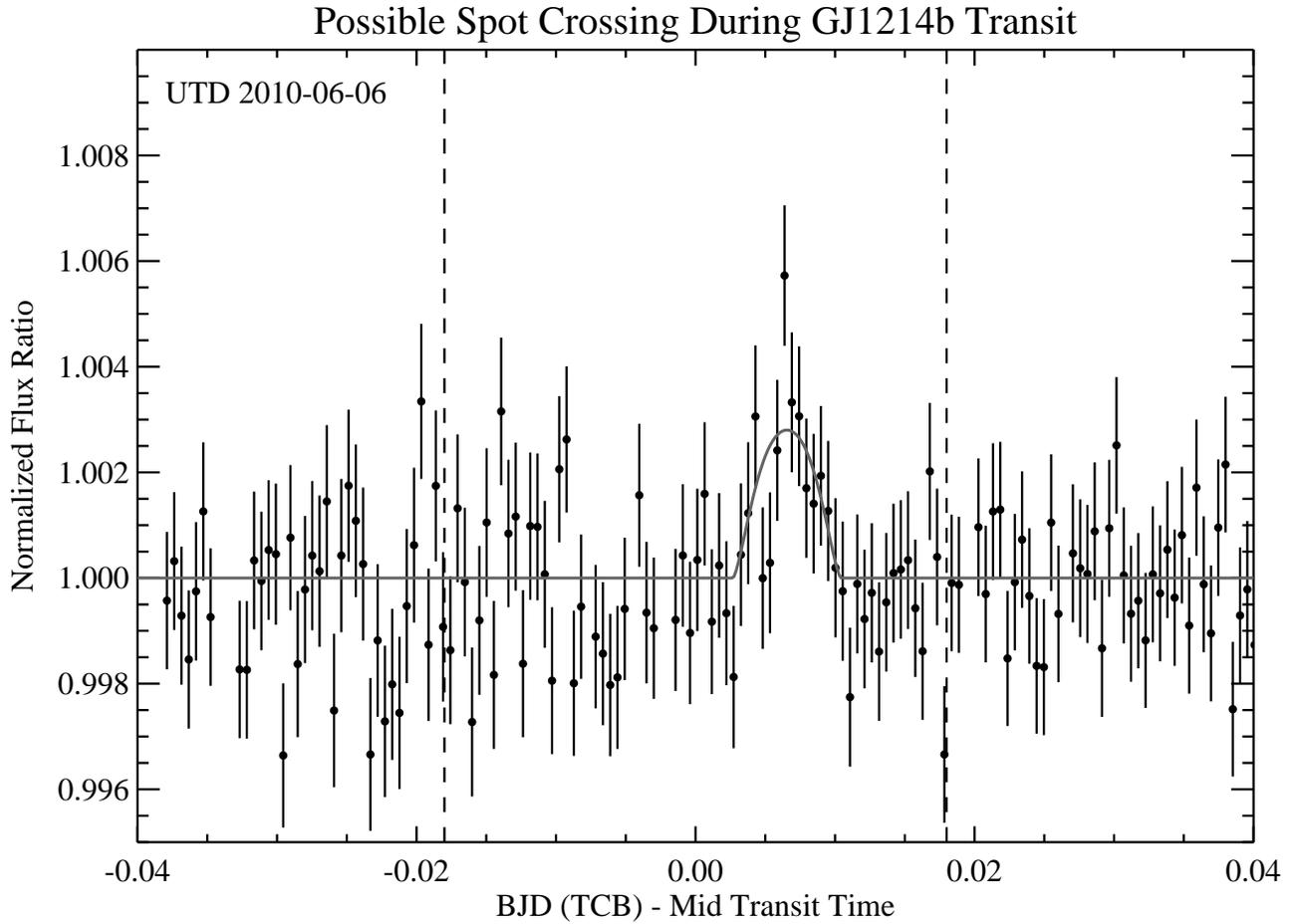}
\caption{\label{figure_spot} Possible spot-crossing event from the UTD 2010-06-06 transit of GJ 1214b. The transit signal was removed by normalizing the lightcurve with the best-fit transit model. The vertical dashed lines approximately mark start and end of the transit. The gray line shows a fit using a simplified spot model.}
\end{figure}

During the transit on UTD 2010-06-06, we observed a slight brightening in the lightcurve signal at the onset of egress (see Figure \ref{figure_lcs}). Figure \ref{figure_spot} shows this event in greater detail. The brightening could be attributed to either another flare event or the planet's temporary obscuration of a cool region on the stellar surface (spot-crossing). The flare model described above provided poor fits to this signal; the shape seen is far more symmetric than the FRED shape of a flare. The symmetry and the fact that it occurred during transit makes it very likely that we observed a spot-crossing event. We modeled the spot-crossing signal based on the analytic expressions in \citet{mandelagol02} for the area of intersection between two circles, assuming the spot was of roughly circular shape. We did not account for the deformation of the spot due to the curvature of the star. We fit for a parameter set very similar to that used for the transit model. Our variables include the spot-ingress duration, the spot-crossing duration, the central crossing time, the square of the radius ratio of the spot to the planet ($R^2_{sp}/R^2_{p}$) and the spot temperature ($T_{sp}$). To account for the fact that the cool region may not be a purely dark spot, we used a contrast ratio dependent on the the stellar and spot temperatures. So the height of the spot signal is roughly equal to $R^2_{sp}/R^2_{p} (1 - I_{sp}/I_{\star})$, where the contrast factor $I_{sp}/I_{\star} = B(T_{sp},\lambda_{obs})/B(T_{\star},\lambda_{obs})$ \citep{silva03}. The function $B(T,\lambda)$ is the Planck function; for this calculation the spot and star were assumed to have blackbody SEDs. The best fit model is shown in gray in Figure \ref{figure_spot}. The model fits the spot feature better than a flat-line, with the relative goodness of fit, $\Delta\chi^2 = 33.3$. We find that a circular spot with a temperature $\sim 2769K$ and spot-to-planet radius ratio of 0.06 fits the feature very well. Using the planetary radius estimated from chain 003a in Table \ref{table_pars2} we estimate the radius of the spot to be $\sim 0.16R_{\earth}$. The duration of spot-crossing ($t_{T,sp}$) can also be used to estimate the longitudinal extent of the cool region, $l_{sp} \propto t_{T,sp}a/$Period $\sim 0.99R_{\earth}$. The hugely different estimates of spot size show just how degenerate the various parameters in our model are. There is not enough information to reconcile degeneracies in the planet-to-spot impact parameter, the spot radius and the contrast ratio between the spot and star.

\section{Conclusions}
\label{sec-conclusions}
\N \textbf{A Transit Model Suited for Bayesian Analysis:} We show that fitting for the transit duration ($t_T$) and the ingress/egress duration ($t_G$) results in a parameter set with few mutual degeneracies (see Figure \ref{figure_mcmc_1}). This condition is suited very well for MCMC methods, which are regularly used to determine uncertainties on parameters derived from transit lightcurves. Our joint analysis of multi-wavelength data using this parameter set was able to reproduce previous estimates of system parameters for GJ 1214b (see Table \ref{table_pars2}, chain003a). We also find that milli-magnitude photometry may not be sufficient to constrain limb-darkening parameters using transit lightcurves. We show that MCMC runs where we fit for these parameters were slow to converge (see Table \ref{table_mcmcstats}), and posterior probability distributions for various parameters were plagued with degeneracies (see Figure \ref{figure_mcmc_2}). Estimates of system parameters from these runs were generally unreliable when compared to runs where the limb-darkening parameters were kept fixed (see chains `b' vs. `a' in Tables \ref{table_pars1} and \ref{table_pars2}).

\N \textbf{Transit Timing Variations:} Data gathered so far do not indicate significant variations in the times of transit for GJ 1214b (see Figure \ref{figure_ttv}). APOSTLE will continue making observations of GJ 1214b and a more detailed analysis of timing data will follow in a future paper.

\N \textbf{System Parameters for GJ 1214:} From fitting SEDs to photometry, we constrained GJ 1214's observed flux and luminosity ($\S$ \ref{sec-mr}). The luminosity allowed us to constrain GJ 1214's mass and since we obtained stellar density from transit lightcurves it allowed us to estimate GJ 1214's radius. We find the derived values of mass and radius to be in agreement with previous estimates, however we find GJ 1214 deviates from well-known mass-radius relations for low-mass stars (see Figure \ref{figure_mrplot}). Simple calculations using the formalism presented in \citet{chabrier07} show that GJ 1214's position on the mass-radius plot can be explained by the presence of cool regions on its surface.

From RV, transit data \citep{charbonneau09} and absolute stellar properties we determined various properties of GJ 1214b (see Table \ref{table_pars3}). The planetary mass and radius (6.37$\pm$0.87 M$_{\earth}$, 2.74$_{-0.05}^{+0.06}$ R$_{\earth}$) places GJ 1214b between the terrestrial and ice-giant regime of planets (2M$_{\earth} <$ M$_{p}$ $<$ 10M$_{\earth}$). Its classification as a ``Super-Earth'' remains and the planetary density confirms it is not like the rocky bodies of our solar system (see Table \ref{table_pars3}). \citet{rogersseager10} propose 3 scenarios for the origin of its gaseous envelope: i) primordial H/He, ii) sublimated ices (H$_2$O,CO$_2$) or iii) volcanic outgassing. \citet{millerrfortney10} propose that space-based observations of the transmission spectra of GJ 1214b's atmosphere should be able to tell us how Hydrogen-rich its atmosphere is. The largest source of uncertainty in our estimate of planetary mass was the velocity semi-amplitude ($K$). Errors in the planetary radius follow from our uncertainty in measuring the absolute size of the star, which ultimately hinges on our luminosity estimate (see $\S$\ref{sec-mr}). Improved precision on radial velocity, flux and distance would tighten our constraints on the absolute mass and radius of GJ 1214b.

\N \textbf{Evidence for Stellar Activity:} The detection of a low energy stellar flare and the possible transit of the planet over a star-spot (see Figures \ref{figure_flare} and \ref{figure_spot}) indicate that GJ 1214 is active. However, considering its age and comparing the flare energy to flares on the younger AD Leo confirms that GJ 1214 is a quiet star for its spectral type \citep{hawley96,hawley03}. We find a fast-rise exponential decay profile fits the flare signal (UTD 2010-04-21) quite well.

A symmetric rise in the normalized flux ratio during the transit on UTD 2010-06-06 could indicate the planet occulted a star-spot on the surface of GJ 1214. The signal is weak, and our attempt at fitting a simplified spot model shows that spot properties are difficult to constrain from a single spot-crossing observation. Our results on spot-properties are inconclusive due to degeneracies between spot-size, planet-to-spot impact parameter and spot-to-star contrast ratio in our model. Detections of this signal from successive transits would have confirmed it as a star-spot and provided interesting constraints on the properties of an active stellar surface region \citep{dittmann09}. The stellar rotation rate might have also been estimated with such data.

\section*{Acknowledgments}
Funding for this work came from NASA Origins grant NNX09AB32G and NSF Career grant 0645416. We would like to thank the APO Staff and Engineers for helping the APOSTLE program with its observations and instrument characterization. We would like thank S. L. Hawley and A. Kowalski for very helpful discussions on M-Dwarfs and stellar flares and J. Davenport for providing useful tips on fast contour plotting in IDL. A large part of our analysis depended on data from \citet{charbonneau09}. We are indebted to the MEarth team for providing us their lightcurves. We are also grateful to J. M. D\'{e}sert for providing us with tentative Warm Spitzer measurements of GJ 1214's flux.


\begin{sidewaystable}[!h]\scriptsize
\caption{\label{table_mcmcstats} MCMC Analysis}
\begin{tabular}{cccrrrrrr}
\hline \hline
Chain Name & Data Set & Model Pars & N$_{pars}$ & Chain Length & Corr Length & Eff Length & $\chi^2$ & DOF \\
\hline \hline
001a & APOSTLE & $\boldsymbol{\theta}_1$ & 6 & 998,301 & 11 & 90,755 & 802 & 808 \\   
001b & APOSTLE & $\boldsymbol{\theta}_1$ & 8 & 1,193,465 & 354 & 3,371 & 796 & 806 \\ 
002a & MEarth + FLWO1.2m & $\boldsymbol{\theta}_2$ & 9 & 998,308 & 66 & 15,126 & 1279 & 1294 \\ 
002b & MEarth + FLWO1.2m & $\boldsymbol{\theta}_2$ & 13 & 1,193,569 & 939 & 1,271 & 1249 & 1290 \\ 
003a & APOSTLE + MEarth + FLWO1.2m & $\boldsymbol{\theta}_2$ & 12 & 997,202 & 84 & 11,871 & 2067 & 2105 \\ 
003b & APOSTLE + MEarth + FLWO1.2m & $\boldsymbol{\theta}_2$ & 18 & 2,398,779 & 3,299 & 727 & 1947 & 2099 \\ 
004a & MEarth & $\boldsymbol{\theta}_1$ & 7 & 996,616 & 16 & 62,289 & 804 & 813 \\ 
004b & MEarth & $\boldsymbol{\theta}_1$ & 9 & 1,195,033 & 240 & 4,979 & 796 & 811\\ 
005a & FLWO1.2m & $\boldsymbol{\theta}_1$ & 5 & 997,093 & 8 & 124,637 & 475 & 478 \\ 
005b & FLWO1.2m & $\boldsymbol{\theta}_1$ & 7 & 1,198,883 & 148 & 8,101 & 471 & 476\\ 
\hline \hline
\end{tabular}
\vspace{-3.2mm}
\end{sidewaystable}

\begin{sidewaystable}\tiny
\begin{center}
\caption{\label{table_pars1} Parameter set $\boldsymbol{\theta}_{1}$}
\begin{tabular}{ccccccccc}
\hline \hline
Parameter/Chain & 001a & 001b & 004a & 004b & 005a & 005b & \citet{charbonneau09}$^{*}$ & Units \\ [2pt]
\hline \hline
\multicolumn{9}{c}{Model} \\
\hline
D& 0.0135 $\pm$ 0.0002& 0.0130$_{-0.0001}^{+0.0007}$& 0.0141 $\pm$ 0.0001& 0.0133$_{-0.0003}^{+0.0001}$& 0.0145 $\pm$ 0.0003& 0.0136$_{-0.0005}^{+0.0003}$& 0.0135 $\pm$ 0.0002& - \\ [2pt]
$t_{T}$& 0.0326 $\pm$ 0.0002& 0.0329$_{-0.0011}^{+0.0001}$& 0.0311 $\pm$ 0.0001& 0.0315$_{-0.0004}^{+0.0003}$& 0.0314 $\pm$ 0.0002& 0.0319$_{-0.0007}^{+0.0006}$& 0.0321 $\pm$ 0.1529& day \\ [2pt]
$t_{G}$& 0.0048$_{-0.0006}^{+0.0002}$& 0.0038$_{-0.0003}^{+0.0014}$& 0.0037$_{-0.0001}^{+0.0009}$& 0.0040$_{-0.0001}^{+0.0006}$& 0.0051$_{-0.0002}^{+0.0009}$& 0.0045$_{-0.0004}^{+0.0010}$& 0.0043 $\pm$ 0.0203& day \\ [2pt]
v1$_{APOSTLE}$& (0.908)& 1.00$_{-0.33}^{+0.06}$& - & - & - & - & - & - \\ [2pt]
v2$_{APOSTLE}$& (0.305)& 0.42$_{-0.07}^{+0.59}$& - & - & - & - & - & - \\ [2pt]
v1$_{M_{Earth}}$& - & - & (0.145)& 0.28 $\pm$ 0.12& - & - & (0.145)& - \\ [2pt]
v2$_{M_{Earth}}$& - & - & (0.639)& 1.24$_{-0.42}^{+0.52}$& - & - & (0.639)& - \\ [2pt]
v1$_{FLWO1.2m}$& - & - & - & - & (0.404)& 0.51$_{-0.26}^{+0.23}$& (0.404)& - \\ [2pt]
v2$_{FLWO1.2m}$& - & - & - & - & (-0.289)& 0.30$_{-0.57}^{+0.93}$& (-0.289)& - \\ [2pt]
T0.T1 2455307+& 0.892740 $\pm$ 0.000275& 0.892769 $\pm$ 0.000276& - & - & - & - & - & BJD$_{TCB}$ \\ [2pt]
T0.T2 2455353+& 0.724696 $\pm$ 0.000317& 0.724738 $\pm$ 0.000317& - & - & - & - & - & BJD$_{TCB}$ \\ [2pt]
T0.T3 2455383+& 0.752366 $\pm$ 0.000265& 0.752306 $\pm$ 0.000264& - & - & - & - & - & BJD$_{TCB}$ \\ [2pt]
T0.T4 2454964+& - & - & 0.944783 $\pm$ 0.000794& 0.944955 $\pm$ 0.000792& - & - & 0.945142 $\pm$ 0.000403& BJD$_{TCB}$ \\ [2pt]
T0.T5 2454980+& - & - & 0.748909 $\pm$ 0.000272& 0.748904 $\pm$ 0.000271& - & - & 0.748903 $\pm$ 0.000090& BJD$_{TCB}$ \\ [2pt]
T0.T6 2454983+& - & - & 0.909659 $\pm$ 0.000233& 0.909675 $\pm$ 0.000233& - & - & 0.909693 $\pm$ 0.000090& BJD$_{TCB}$ \\ [2pt]
T0.T7 2454999+& - & - & 0.713612 $\pm$ 0.000260& 0.713654 $\pm$ 0.000260& - & - & 0.713634 $\pm$ 0.000126& BJD$_{TCB}$ \\ [2pt]
T0.T8 2454980+& - & - & - & - & 0.748974 $\pm$ 0.000467& 0.748886 $\pm$ 0.000469& - & BJD$_{TCB}$ \\ [2pt]
T0.T9 2454983+& - & - & - & - & 0.909754 $\pm$ 0.000444& 0.909699 $\pm$ 0.000446& - & BJD$_{TCB}$ \\ [2pt]
\hline
\multicolumn{9}{c}{Derived} \\
\hline
$R_{p}/R_{\star}$& 0.1234$_{-0.0047}^{+0.0008}$& 0.1143$_{-0.0033}^{+0.0134}$& 0.1187$_{-0.0006}^{+0.0042}$& 0.1182$_{-0.0015}^{+0.0048}$& 0.1221$_{-0.0010}^{+0.0020}$& 0.1205$_{-0.0034}^{+0.0068}$& 0.1162 $\pm$ 0.0007& - \\ [2pt]
P(1.5804 days+)& 0.79 $\pm$ 0.25& 0.65 $\pm$ 0.25& 0.10 $\pm$ 1.38& -0.40 $\pm$ 1.39& -0.86 $\pm$ 9.97& 0.52 $\pm$ 9.91& -0.65 $\pm$ 1.01&sec\\ [2pt]
b& 0.41$_{-0.12}^{+0.03}$& 0.11$_{-0.15}^{+0.35}$& 0.04$_{-0.14}^{+0.39}$& 0.25$_{-0.06}^{+0.14}$& 0.50$_{-0.03}^{+0.10}$& 0.38$_{-0.10}^{+0.14}$& 0.35$_{-0.08}^{+0.06}$& - \\ [2pt]
a/$R_{\star}$& 14.06$_{-0.19}^{+0.72}$& 15.18$_{-1.26}^{+0.27}$& 16.19$_{-1.46}^{+0.21}$& 15.47$_{-0.67}^{+0.19}$& 13.90$_{-1.03}^{+0.26}$& 14.61$_{-0.95}^{+0.43}$& 14.66 $\pm$ 0.41& - \\ [2pt]
i& 88.32$_{-0.13}^{+0.56}$& 89.59$_{-1.49}^{+0.57}$& 89.87$_{-1.53}^{+0.49}$& 89.08$_{-0.58}^{+0.24}$& 87.94$_{-0.61}^{+0.17}$& 88.52$_{-0.68}^{+0.44}$& 88.62$_{-0.28}^{+0.35}$& deg\\ [2pt]
$v$& 55.90$_{-0.74}^{+2.85}$& 60.34$_{-5.02}^{+1.07}$& 64.35$_{-5.79}^{+0.85}$& 61.52$_{-2.67}^{+0.75}$& 55.25$_{-4.09}^{+1.02}$& 58.07$_{-3.78}^{+1.71}$& 58.28 $\pm$ 1.63&day$^{-1}$\\ [2pt]
$\rho_{\star}$& 21.06$_{-0.82}^{+3.39}$& 26.49$_{-6.08}^{+1.38}$& 32.13$_{-7.91}^{+1.26}$& 28.06$_{-3.50}^{+1.04}$& 20.34$_{-4.19}^{+1.15}$& 23.61$_{-4.31}^{+2.15}$& 23.90 $\pm$ 2.10&g/cm$^{3}$\\ [2pt]
\hline \hline
\end{tabular}
\vspace{-3.2mm}
\end{center}
$*$ For those parameters not explicitly listed by \citet{charbonneau09} (e.g. t$_{T}$ and t$_{G}$) we computed them using the expressions in \citet{carter08} and propagated the errors assuming they had Gaussian distributions.
\end{sidewaystable}

\begin{sidewaystable}\tiny
\begin{center}
\caption{\label{table_pars2} Parameter set $\boldsymbol{\theta}_{2}$}
\begin{tabular}{ccccccc}
\hline \hline
Parameter/Chain & 002a & 002b & 003a & 003b & \citet{charbonneau09}$^{*}$ & Units \\ [2pt]
\hline \hline
\multicolumn{7}{c}{Model} \\
\hline
$R_{p}^2/R_{\star}^2$& 0.0145 $\pm$ 0.0002& 0.0137$_{-0.0002}^{+0.0007}$& 0.0143$_{-0.0001}^{+0.0002}$& 0.0133$_{-0.0001}^{+0.0006}$& 0.0135 $\pm$ 0.0002& - \\ [2pt]
$t_{T}$& 0.0310$_{-0.0002}^{+0.0001}$& 0.0316$_{-0.0004}^{+0.0002}$& 0.0316 $\pm$ 0.0001& 0.0321$_{-0.0004}^{+0.0002}$& 0.0321 $\pm$ 0.1529& day \\ [2pt]
$t_{G}$& 0.0045$_{-0.0003}^{+0.0005}$& 0.0041$_{-0.0002}^{+0.0006}$& 0.0045$_{-0.0001}^{+0.0002}$& 0.0037 $\pm$ 0.0002& 0.0043 $\pm$ 0.0203& day \\ [2pt]
v1$_{APOSTLE}$& - & - & (0.908)& 0.71$_{-0.08}^{+0.14}$& - & - \\ [2pt]
v2$_{APOSTLE}$& - & - & (0.305)& 1.01$_{-0.43}^{+0.20}$& - & - \\ [2pt]
v1$_{M_{Earth}}$& (0.145)& 0.38$_{-0.21}^{+0.08}$& (0.145)& 0.53$_{-0.21}^{+0.11}$& (0.145)& - \\ [2pt]
v2$_{M_{Earth}}$& (0.639)& 1.05$_{-0.31}^{+0.72}$& (0.639)& 0.47$_{-0.28}^{+0.68}$& (0.639)& - \\ [2pt]
v1$_{FLWO1.2m}$& (0.404)& 0.45$_{-0.19}^{+0.15}$& (0.404)& 0.61$_{-0.20}^{+0.13}$& (0.404)& - \\ [2pt]
v2$_{FLWO1.2m}$& (-0.289)& 0.55 $\pm$ 0.58& (-0.289)& 0.24 $\pm$ 0.51& (-0.289)& - \\ [2pt]
T0.T1 2455307+& - & - & 0.892642 $\pm$ 0.000271& 0.892689 $\pm$ 0.000263& - & BJD$_{TCB}$ \\ [2pt]
T0.T2 2455353+& - & - & 0.724728 $\pm$ 0.000307& 0.724652 $\pm$ 0.000311& - & BJD$_{TCB}$ \\ [2pt]
T0.T3 2455383+& - & - & 0.752332 $\pm$ 0.000260& 0.752334 $\pm$ 0.000264& - & BJD$_{TCB}$ \\ [2pt]
T0.T4 2454964+& 0.944943 $\pm$ 0.000794& 0.944940 $\pm$ 0.000789& 0.944962 $\pm$ 0.000799& 0.944935 $\pm$ 0.000788& 0.945142 $\pm$ 0.000403& BJD$_{TCB}$ \\ [2pt]
T0.T5 2454980+& 0.748921 $\pm$ 0.000266& 0.748938 $\pm$ 0.000267& 0.748905 $\pm$ 0.000263& 0.748976 $\pm$ 0.000264& 0.748903 $\pm$ 0.000090& BJD$_{TCB}$ \\ [2pt]
T0.T6 2454983+& 0.909665 $\pm$ 0.000229& 0.909650 $\pm$ 0.000230& 0.909674 $\pm$ 0.000228& 0.909689 $\pm$ 0.000228& 0.909693 $\pm$ 0.000090& BJD$_{TCB}$ \\ [2pt]
T0.T7 2454999+& 0.713648 $\pm$ 0.000256& 0.713675 $\pm$ 0.000257& 0.713625 $\pm$ 0.000253& 0.713690 $\pm$ 0.000253& 0.713634 $\pm$ 0.000126& BJD$_{TCB}$ \\ [2pt]
T0.T8 2454980+& 0.748938 $\pm$ 0.000380& 0.748897 $\pm$ 0.000414& 0.748924 $\pm$ 0.000390& 0.748942 $\pm$ 0.000417& - & BJD$_{TCB}$ \\ [2pt]
T0.T9 2454983+& 0.909734 $\pm$ 0.000373& 0.909715 $\pm$ 0.000400& 0.909736 $\pm$ 0.000381& 0.909686 $\pm$ 0.000401& - & BJD$_{TCB}$ \\ [2pt]
\hline
\multicolumn{7}{c}{Derived} \\
\hline
$R_{p}/R_{\star}$& 0.1203$_{-0.0006}^{+0.0009}$& 0.1171$_{-0.0008}^{+0.0029}$& 0.1195$_{-0.0004}^{+0.0008}$& 0.1152$_{-0.0003}^{+0.0026}$& 0.1162 $\pm$ 0.0007& - \\ [2pt]
P(1.5804 days+)& -0.39 $\pm$ 1.38& -0.27 $\pm$ 1.42& 0.47 $\pm$ 0.04& 0.46 $\pm$ 0.04& -0.65 $\pm$ 1.01&sec\\ [2pt]
b& 0.42$_{-0.07}^{+0.08}$& 0.32$_{-0.07}^{+0.13}$& 0.41$_{-0.03}^{+0.04}$& 0.09$_{-0.04}^{+0.24}$& 0.35$_{-0.08}^{+0.06}$& - \\ [2pt]
a/$R_{\star}$& 14.70$_{-0.67}^{+0.49}$& 15.08$_{-0.74}^{+0.30}$& 14.54$_{-0.33}^{+0.19}$& 15.60$_{-0.72}^{+0.08}$& 14.66 $\pm$ 0.41& - \\ [2pt]
i& 88.36$_{-0.42}^{+0.34}$& 88.79$_{-0.58}^{+0.29}$& 88.39$_{-0.21}^{+0.14}$& 89.67$_{-0.94}^{+0.15}$& 88.62$_{-0.28}^{+0.35}$& deg\\ [2pt]
$v$& 58.46$_{-2.66}^{+1.96}$& 59.94$_{-2.95}^{+1.19}$& 57.81$_{-1.32}^{+0.75}$& 62.02$_{-2.87}^{+0.33}$& 58.28 $\pm$ 1.63&day$^{-1}$\\ [2pt]
$\rho_{\star}$& 24.09$_{-3.14}^{+2.50}$& 25.97$_{-3.65}^{+1.58}$& 23.29$_{-1.55}^{+0.92}$& 28.76$_{-3.82}^{+0.45}$& 23.90 $\pm$ 2.10&g/cm$^{3}$\\ [2pt]
\hline \hline
\end{tabular}
\vspace{-3.2mm}
\end{center}
$*$ For those parameters not explicitly listed by \citet{charbonneau09} (e.g. t$_{T}$ and t$_{G}$) we computed them using the expressions in \citet{carter08} and propagated the errors assuming they had Gaussian distributions.
\end{sidewaystable}

\begin{sidewaystable}
\caption{\label{table_pars3}Properties of GJ 1214 and GJ 1214b}
\begin{tabular}{lll}
\hline \hline
Parameter & Value & Units \\
\hline
\multicolumn{3}{c}{Fit SED Parameters} \\
\hline
$T_{eff}$& 2949$_{-32}^{+27}$&K\\
$\log{g}$& 4.94$_{-0.26}^{+0.22}$&$\log{cm/s^2}$\\
$R_{\star}/d$& 0.0155 $\pm$ 0.0003&$R_{\sun}/pc$\\
\hline
\multicolumn{3}{c}{Derived Stellar Parameters} \\
\hline
$F_{obs}$& 5.23 $\pm$ 0.13&$10^{-10} ergs/s/cm^{2}$\\
$L_{\star}$& 0.0028 $\pm$ 0.0004&$L_{\sun}$\\
$M_{\star}$& 0.153$_{-0.009}^{+0.010}$&$M_{\sun}$\\
$R_{\star}$& 0.210$_{-0.004}^{+0.005}$&$R_{\sun}$\\
\hline
\multicolumn{3}{c}{Derived Planetary Parameters} \\
\hline
$M_{p}$& 6.37 $\pm$ 0.87&$M_{\earth}$\\
$R_{p}$& 2.74$_{-0.05}^{+0.06}$&$R_{\earth}$\\
$\rho_{p}$& 1.68 $\pm$ 0.23&$g/cm^3$\\
$g_{p}$& 8.24 $\pm$ 1.09&$m/s^2$\\
$V_{esc,p}$& 12.03 $\pm$ 0.80&km/s\\
$T_{eq}$ (Bond albedo = 0)& 547$_{-8}^{+7}$&K\\
$T_{eq}$ (Bond albedo = 0.75)& 387$_{-6}^{+5}$&K\\
\hline \hline
\end{tabular}
\vspace{-3.2mm}
\end{sidewaystable}

\end{document}